\documentclass{aa}
\usepackage{hyperref}

\usepackage{times}

\usepackage{graphicx}

\usepackage{amssymb}

\def\zz{\sf}

\def\psh{PSR B0656+14}    

\def\smallerfont{\footnotesize}
\def\beq{\begin{equation}}
\def\eeq{\end{equation}}
\def\etal{ et al. }
\def\hst{{\sl HST}}

\def\twolines#1#2{
\renewcommand{\arraystretch}{0.8}
\begin{tabular}{@{}l@{}}
#1 \vrule height3.2ex width0ex \\ #2 \\[.7ex]
\end{tabular}
}

\def\widerul{\vrule height 2.5ex width 0ex depth 0ex}

\def\m@th{\mathsurround=0pt }  
\def\eqalign#1{\null\,\vcenter{\openup1\jot \m@th 
   \ialign{\strut\hfil$\displaystyle{##}$&$\displaystyle{{}##}$\hfil 
   \crcr#1\crcr}}\,} 

\begin{document}

\thesaurus{06.                  
evolution of stars
                (08.14.1;       
                08.16.7;        
                }

\title{
Optical photometry of the PSR B0656+14 \\ and its neighborhood
}

\author{
     A.B.~Koptsevich\inst{1,2,3}
\and G.G.~Pavlov\inst{2}
\and S.V.~Zharikov\inst{3,4}
\and V.V.~Sokolov\inst{3}
\and Yu.A.~Shibanov\inst{1}
\and V.G.~Kurt\inst{5}
}

\offprints{A.B.~Koptsevich}

\institute{
Ioffe Physical Technical Institute, Politekhnicheskaya 26,
St. Petersburg, 194021, Russia \\
\href{mailto:kopts@astro.ioffe.rssi.ru}{kopts@astro.ioffe.rssi.ru},
\href{mailto:shib@stella.ioffe.rssi.ru}{shib@stella.ioffe.rssi.ru}
\and
The Pennsylvania State University,
Dept. of Astronomy \& Astrophysics, 525 Davey Lab,
University Park, PA 16802, USA \\
\href{mailto:pavlov@astro.psu.edu}{pavlov@astro.psu.edu}
\and
Special Astrophysical Observatory of RAS,
Karachai-Cherkessia, Nizhnij Arkhyz, 357147, Russia \\
\href{mailto:sokolov@sao.ru}{sokolov@sao.ru}
\and
Observatorio Astronomico Nacional de Instituto de Astronomia de UNAM, Ensenada,  B.C., 22860, Mexico \\
\href{zhar@astrosen.unam.mx}{zhar@astrosen.unam.mx}
\and
Astro Space Center of the Russian Academy of Sciences, Moscow, 117810, Russia \\
\href{mailto:vkurt@asc.rssi.ru}{vkurt@asc.rssi.ru}
}

\date{}

\authorrunning{Koptsevich et al.}

\maketitle


\begin{abstract}

We present the results of the broad-band photometry
of the nearby middle-aged radio pulsar PSR B0656+14 and its neighborhood
obtained with the 6-meter telescope of the SAO RAS and with the {\sl Hubble
Space Telescope}.
The broad-band
spectral flux $F_\nu$
of the pulsar decreases with increasing frequency in the near-IR
range and increases with frequency in the near-UV range.
The increase towards UV can
be naturally interpreted as
the Rayleigh-Jeans tail of the soft thermal
component of the X-ray spectrum emitted
from the 
surface of the cooling neutron star.
Continuation of the power-law component, which dominates  
in the
high-energy tail of the
X-ray spectrum, to the IR-optical-UV frequencies is consistent with the
observed fluxes. 
This suggests that
the non-thermal pulsar radiation may be of the same origin 
in a broad frequency range from IR to hard X-rays.
We also studied 4 objects detected in the pulsar's
5\arcsec\ neighborhood.

\end{abstract}


\section{Introduction}

\label{s:introduction}

The radio pulsar \object{PSR B0656+14} was discovered by \cite{m78}.
Its basic parameters are presented in Table~\ref{t:basic}.
This pulsar has been detected in X-rays,
at $E\simeq 0.1$--10 keV,
with the space observatories
{\sl Einstein} (\cite{chmm89}),
{\sl ROSAT} (\cite{fok92}; \cite{acprt93}),
{\sl EUVE} (\cite{foe94}; \cite{feb96}),
and {\sl ASCA} (\cite{gcfmos}; \cite{zph00}).
Together with Geminga, PSR B1055$-$52, and, perhaps, the Vela pulsar, 
it 
belongs to a group of middle-aged pulsars 
with thermal-like soft X-ray spectra 
(\cite{BeckTr_NS}, and references therein). 
The X-ray radiation of \psh\ can be described 
as a combination of radiation 
from the entire surface of a cooling neutron 
star (NS) and radiation 
from hotter regions, presumably
polar caps heated by 
relativistic particles produced 
in the pulsar magnetosphere. 
An excess over the hot thermal 
component at energies above $\approx 2$ keV was interpreted 
as nonthermal radiation from pulsar's magnetosphere
(\cite{gcfmos}).
The pulsar has also been marginally detected in gamma-rays
($E > 50$ MeV), at a 3$\sigma$ level (\cite{r96}).

The optical counterpart of \psh\/ was 
detected 
by \cite{Carav_NTT} with the ESO 3.6-meter and NTT telescopes 
in the V band (V$\approx 25$). It was observed 
in the long-pass filter F130LP ($\lambda\!\lambda$ = 2310--4530~\AA) 
with the Faint Object Camera (FOC) aboard the {\sl Hubble Space Telescope}
(\hst) by Pavlov \etal\ (1996; hereafter \cite{PSC}),
who showed that the bulk of radiation
in this band is of a nonthermal origin. 
Pavlov \etal\ (1997; hereafter \cite{PWC}) 
measured the pulsar's fluxes in narrower FOC bands,
F430W, F342W and F195W, and estimated thermal and nonthermal
contributions to the optical-UV flux.
\cite{Mign} measured the flux in the F555W filter of
the \hst\ Wide Field Planetary Camera (WFPC2).
Coherent optical pulsations 
with the radio pulsar period have been detected in the B-band 
by \cite{Shear}.
Kurt et al.~(1998; hereafter \cite{Kurt}) 
presented first results of
the ground-based BVRI photometry of the pulsar
with the 6-meter Big Alt-azimuth Telescope (BTA)
of the Special Astrophysical Observatory
of the Russian Academy of Sciences (SAO RAS).
 
\begin{table*}[t]
\caption{
Parameters of \psh\ (\cite{Taylor}). }
\label{t:656_prop}
\begin{tabular}{cccccccccccc}
\hline\hline
\multicolumn{6}{c}{Observed}&&\multicolumn{5}{c}{Derived} \widerul\\
\cline{1-7}\cline{9-12}
$P$        & $\dot P$   & $D\!M$       & $l$   & $b$ & $\mu_\alpha^a$  & $\mu_\delta^a$  && $\tau$ & $B$                   & $\dot E$              & $d^b$     \widerul \\
ms         & $10^{-14}$ & cm$^{-3}$ pc & deg   & deg & mas yr$^{-1}$   & mas yr$^{-1}$   && Myr    & G                     & erg $s^{-1}$          & kpc       \\
\hline
384.87     & 5.50       & 14.0         & 201.1 & 8.3 & $43\pm2$        & $-2\pm3$        && 0.11   & $4.7 \times 10^{12}$  & $3.8 \times 10^{34}$  & 0.2--0.8 \widerul\\
\hline
\end{tabular}
\begin{tabular}{ll}
$^a$Pulsar proper motion (\cite{mPM}). \\
$^b$Estimates obtained from interpretations of
 X-ray data give $d=0.2$--0.5 kpc;
the dispersion-measure distance is $\approx 0.76$ kpc. & \\ 
\end{tabular}
\label{t:basic}
\end{table*}

The optical-UV data obtained in the above-described observations  
have shown that the pulsar's optical flux 
exceeds the Rayleigh-Jeans extrapolation of  
the thermal spectrum  seen in soft X-rays. 
\cite{PWC} fit the observed optical-UV fluxes
with a two-component model, a sum of the power-law and Rayleigh-Jeans
spectra.
The slope of the nonthermal component $\alpha_{\rm phot}=-2.4\pm0.7$, 
although poorly constrained in these fits, looks 
steeper than that in X-rays.
The thermal component is characterized by the Rayleigh-Jeans
parameter, $G\equiv T_6 (R_{10}/d_{500})^2 = 1$--6, 
where $T_6$ is the brightness
temperature in $10^6$ K, $R=10\, R_{10}$~km the apparent NS radius,
and $d=500\, d_{500}$~pc the distance. To tighten the constraints
on the parameters of the thermal and nonthermal components,
the fluxes should be measured in a broader wavelength range.

The pulsar field has been recently observed with 
the \hst\ Near Infrared Camera and Multi-Object Spectrometer (NICMOS)
in three broad bands, F110W, F160W, and F187W (Harlow et al.~1998).
The pulsar was detected in all the three bands, but the near-IR fluxes
appeared to be significantly lower than the extrapolation of
the BVRI fluxes measured by \cite{Kurt}.

\strut\cite{Kurt}~also detected four unresolved, faint
($V > 25$) optical objects
within $5\arcsec$ from the pulsar counterpart.
Because of low spatial resolution
of the ground-based observations,
it was impossible to determine whether they are just background
objects (point-like or extended)
or some structures associated with the pulsar.

In this paper we report the results of new optical observations of 
\psh\ and the objects in its nearest neighborhood taken with BTA.
We incorporate the available \hst\ NICMOS, 
WFPC2, and FOC observations\footnote{Observations with the NASA/ESA {\sl Hubble Space
Telescope} were obtained at the Space Telescope Science Institute, which
is operated by the Association of Universities for Research in Astronomy,
Incorporated, under NASA contract NAS5-26555.},   
analyze them together with the new and old BTA data, and obtain 
new estimates of the pulsar magnitudes in the BVRI and near-IR bands. 
We also discuss 
the morphology, color spectra and possible nature
of the objects in the pulsar vicinity.  
 
The observations and data reduction are described in Sect.~\ref{s:observations}. 
Multicolor photometry of the pulsar and the objects in its vicinity  
is presented in Sect.~\ref{s:photometry}. The broad-band spectra and 
the nature of the objects 
are discussed in Sect.~\ref{s:discussion}.

\begin{table*}[t]
\smallerfont
\caption{Log of the \psh\ observations with the BTA.}
\label{t:obs}
\vbox{\hbox{
\begin{tabular}{c|c|cccccc}
\hline\hline
Band   & Date   & $t_{\rm exp}^b$ & $t_{\rm obs}^c$ & $Z^d$   & Seeing  & Sky    & Sky \widerul \\
       &              & s          & UT             & deg     & arcsec  & DN s$^{-1}$ pix$^{-1}$ & mag arcsec$^{-2}$ \\
\hline 
B      & 11.11.96$^a$ & 2400       & -              & 34.30   & 1.7     &  2.47  
     & 21.96 \widerul \\
\hline
V      & 11.11.96$^a$ & 2400       & -              & 33.76   & 1.5     &  6.43  
     & 21.20 \widerul \\
\hline
R      & 11.11.96$^a$ & 3000       & -              & 33.82   & 1.5     & 18.17  
     & 20.44 \widerul \\
\cline{2-8}
       & 26.11.97     & 500        & 00:49          & 31.82   & 1.6     & 20.91  
     & 20.30 \\
       &              & 500        & 01:11          & 33.92   & 1.6     & 21.70  
     & 20.25 \\
\cline{2-8}
       & Sum$^e$     & 4000       & -              & 33.59   & 1.6     & 18.96  
     & 20.40 \\
\hline
I      & 11.11.96$^a$ & 3300       & -              & 33.65   & 1.4     & 29.93  
     & 19.04 \widerul     \\
\cline{2-8}
       & 26.11.97     & 600        & 22:37          & 33.56   & 1.5     & 29.32  
     & 19.07 \\
       &              & 600        & 22:49          & 32.42   & 1.5     & 28.68  
     & 19.09 \\
       &              & 600        & 00:37          & 30.90   & 1.6     & 30.46  
     & 19.03 \\
\cline{2-8}
       & Sum$^e$      & 5100       & -              & 33.14   & 1.5     & 29.77  
     & 19.05 \\
\hline
\end{tabular}
}
\hbox{
\begin{tabular}{clccl}
$^a$ & Summed data sets      && $^d$ & Zenith distance  \\  
     & of 1996 (\cite{Kurt}) && $^e$ & Summed R and I exposures \\
$^b$ & Exposure duration     &&      & of 1996 and 1997 \\
$^c$ & Starting UT of the exposure \\
\end{tabular} 
}}
\end{table*}

\begin{figure*}[p]
\setlength{\unitlength}{1mm}
\resizebox{12cm}{!}{
\begin{picture}(105,210)(0,0)
\put ( 0,4)   {\includegraphics[width=45mm,bb=70 303 522
755,clip]{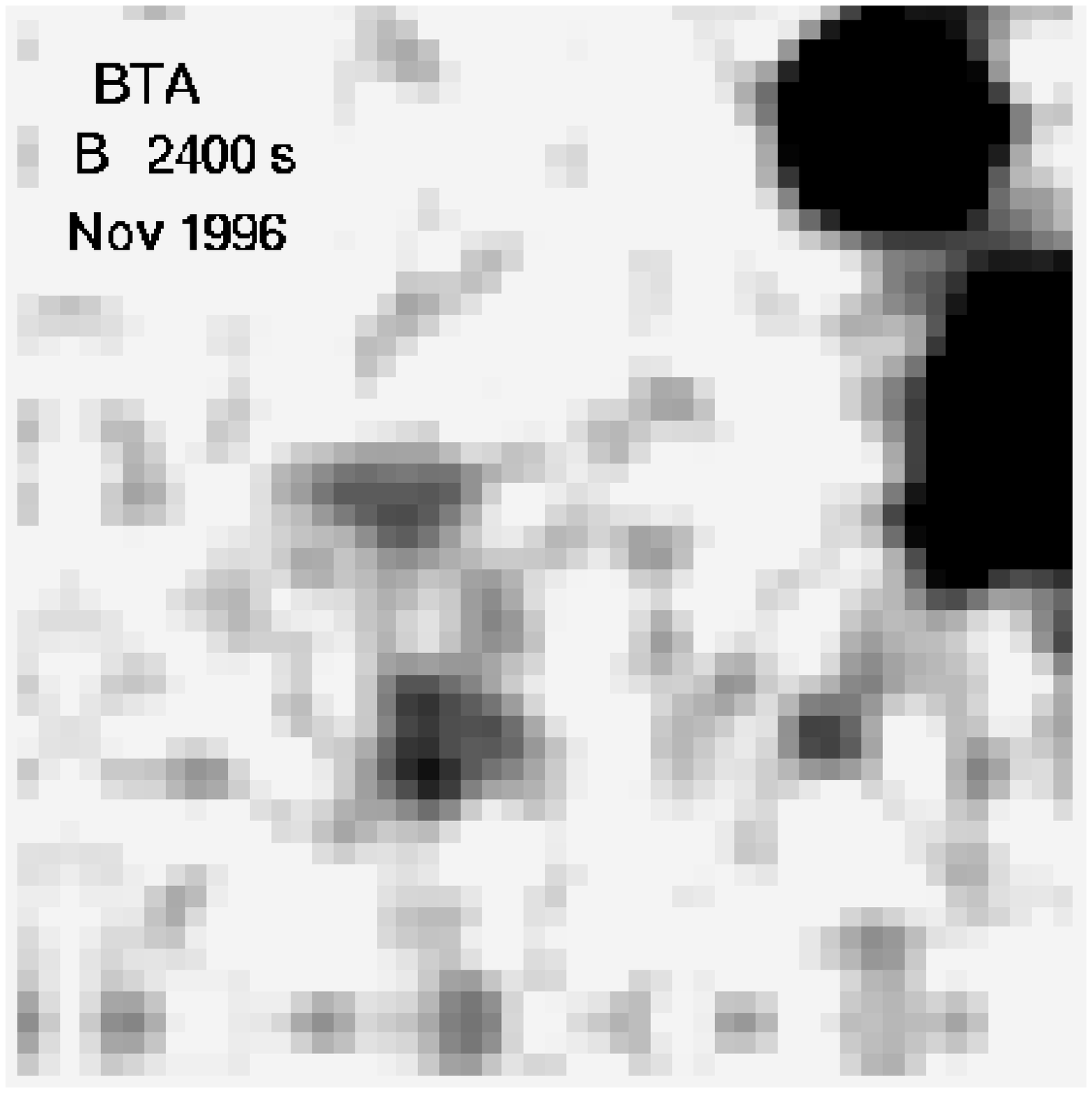}}
\put ( 0,56)   {\includegraphics[width=45mm,bb=70 303 522 
755,clip]{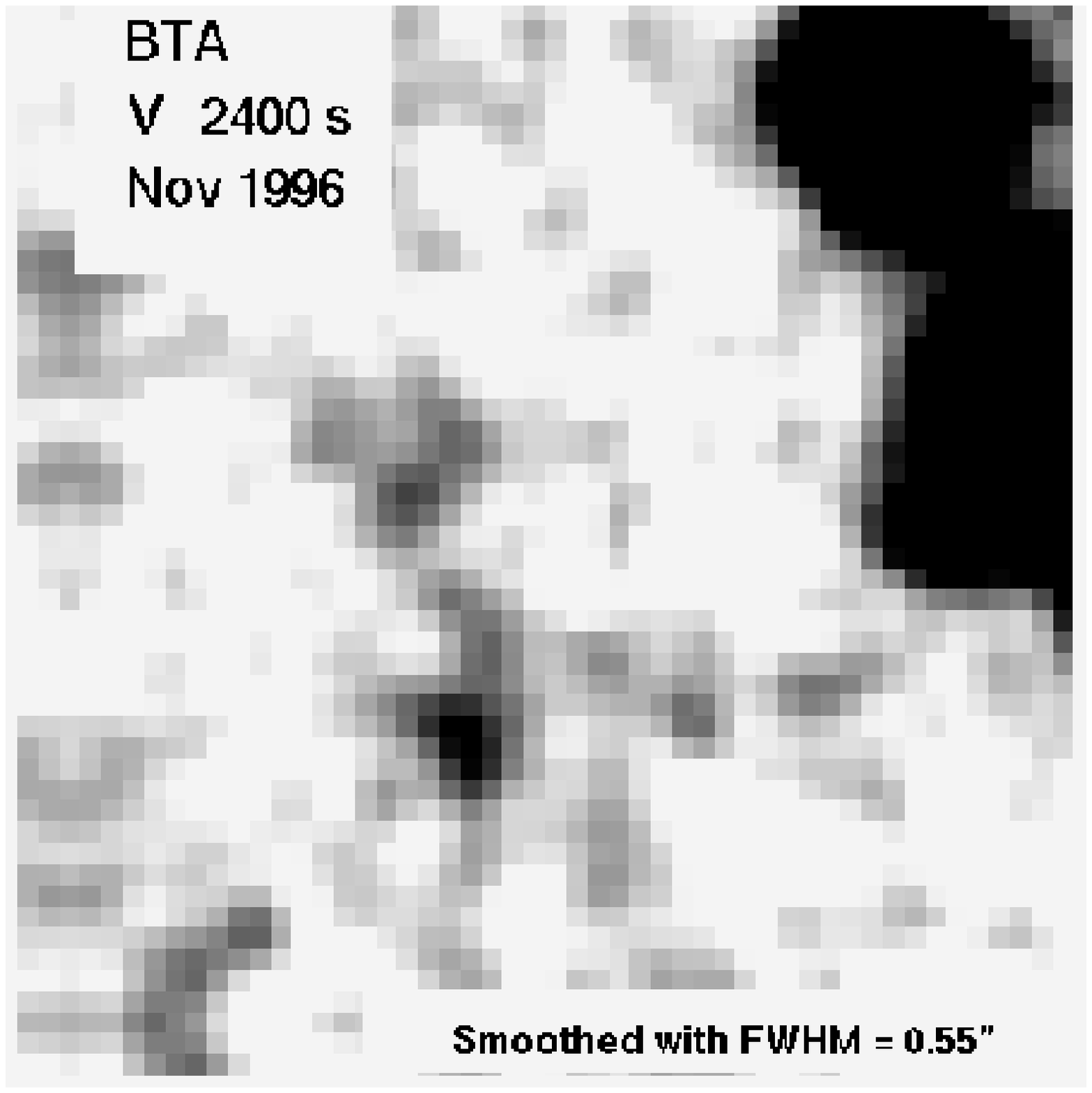}}
\put ( 0,108)   {\includegraphics[width=45mm,bb=70 303 522 
755,clip]{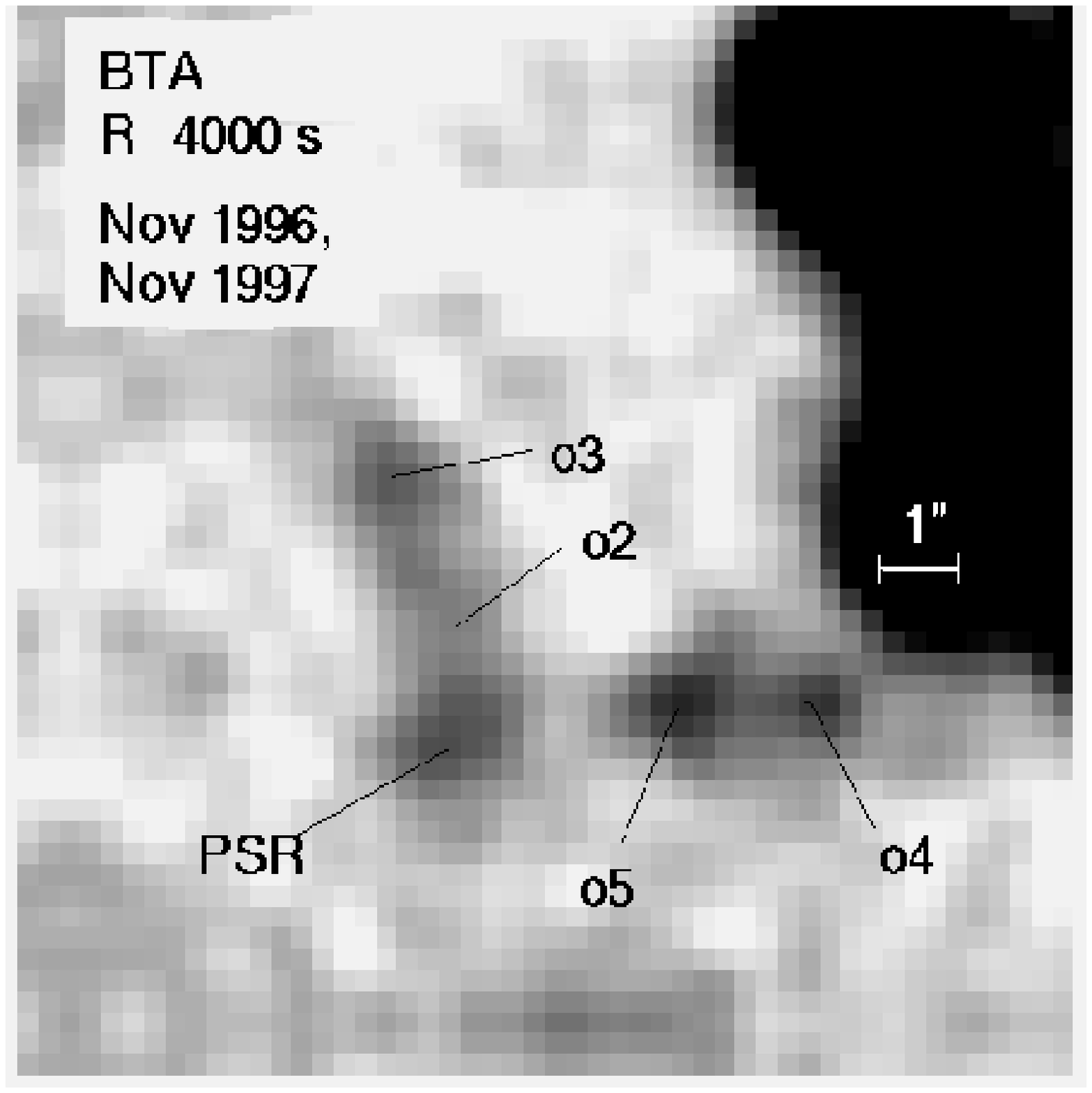}}
\put ( 0,160)   {\includegraphics[width=45mm,bb=70 303 522 
755,clip]{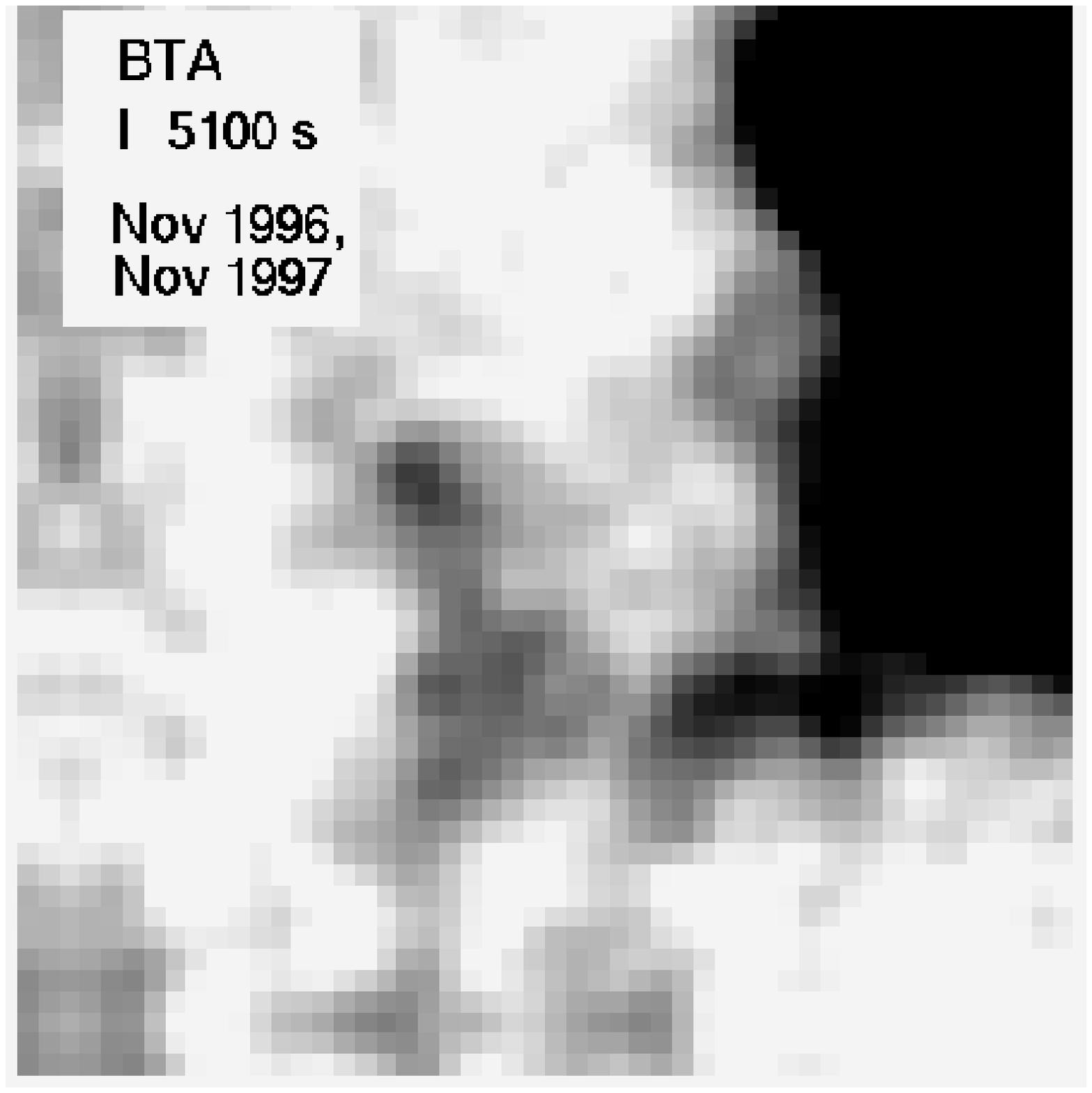}}
\put (50, 0) {\includegraphics[width=50.6mm,height=50.3mm,bb=80 320 540 
780,clip]{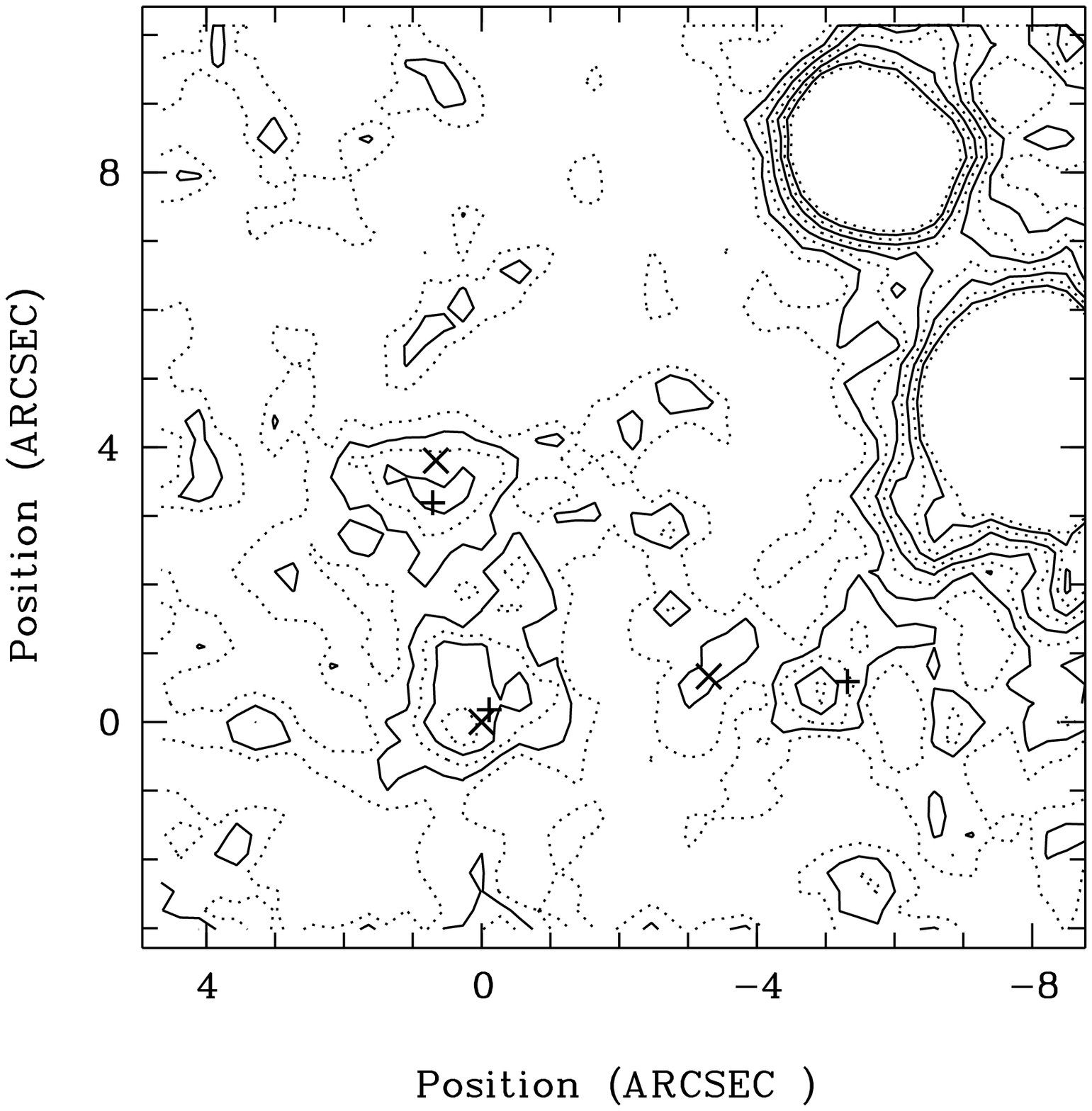}}
\put (50,52) {\includegraphics[width=50.6mm,height=50.3mm,bb=80 320 540 
780,clip]{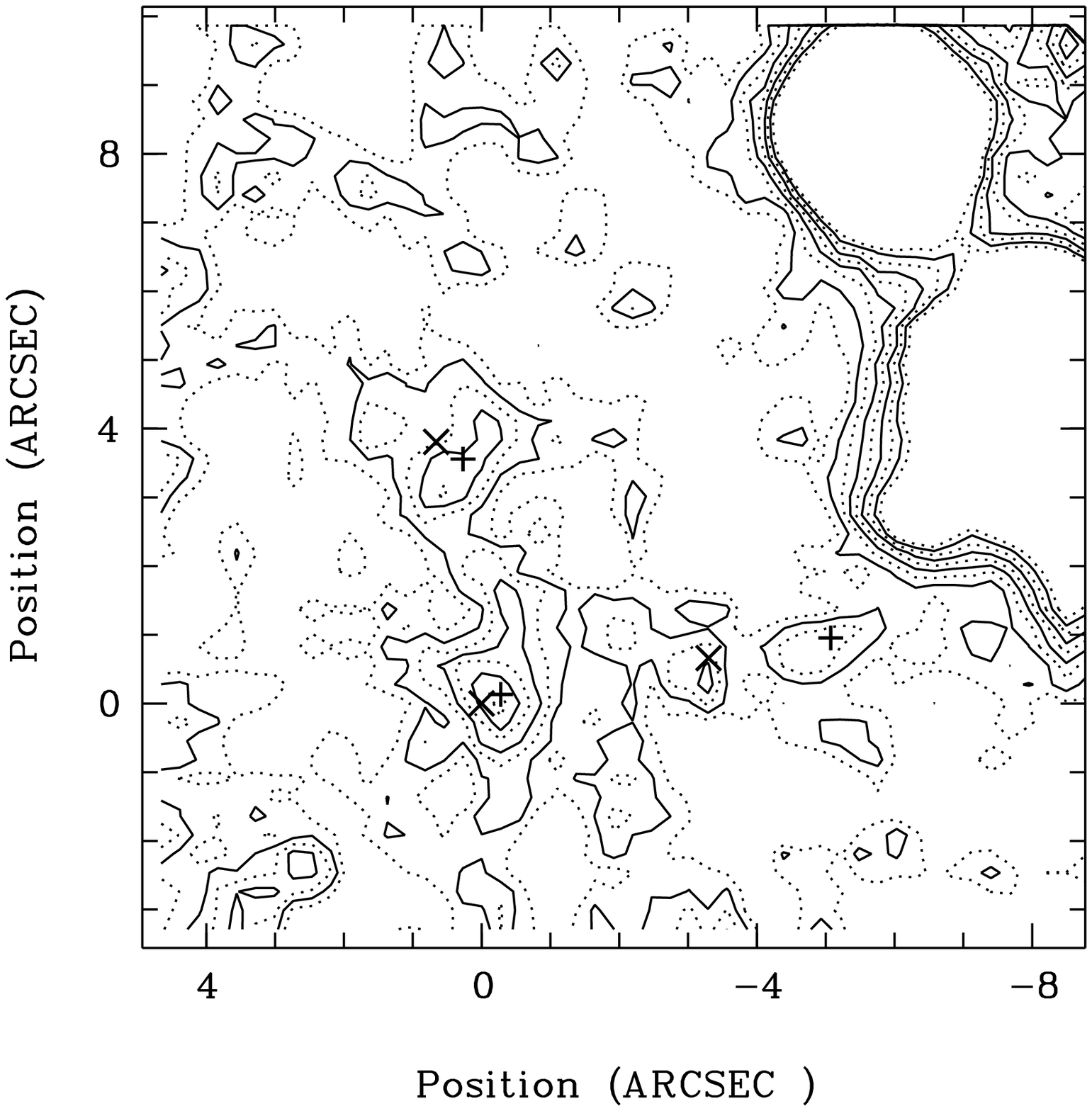}}
\put (50,104) {\includegraphics[width=50.6mm,height=50.3mm,bb=80 320 540 
780,clip]{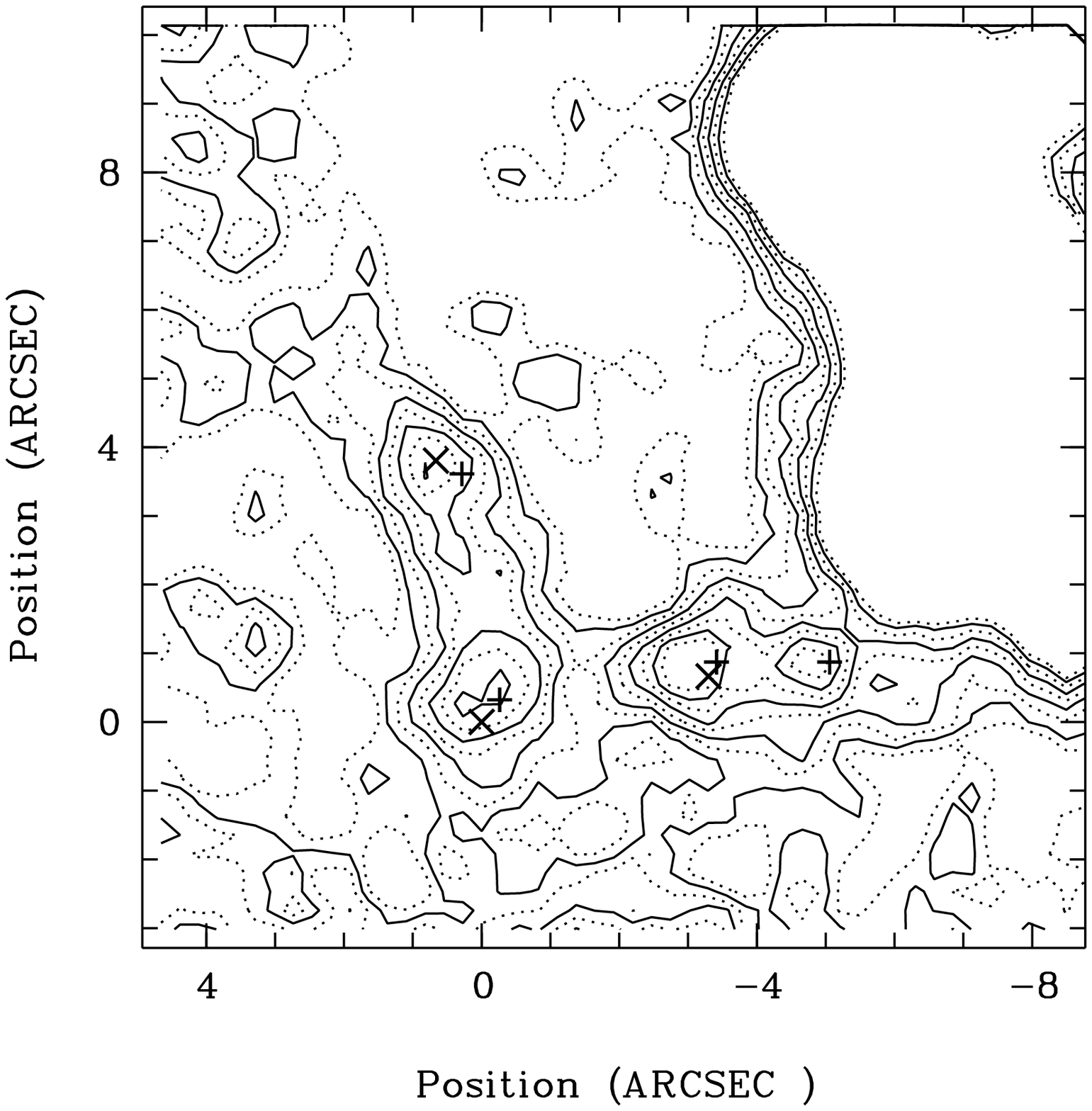}}
\put (50,156) {\includegraphics[width=50.6mm,height=50.3mm,bb=80 320 540 
780,clip]{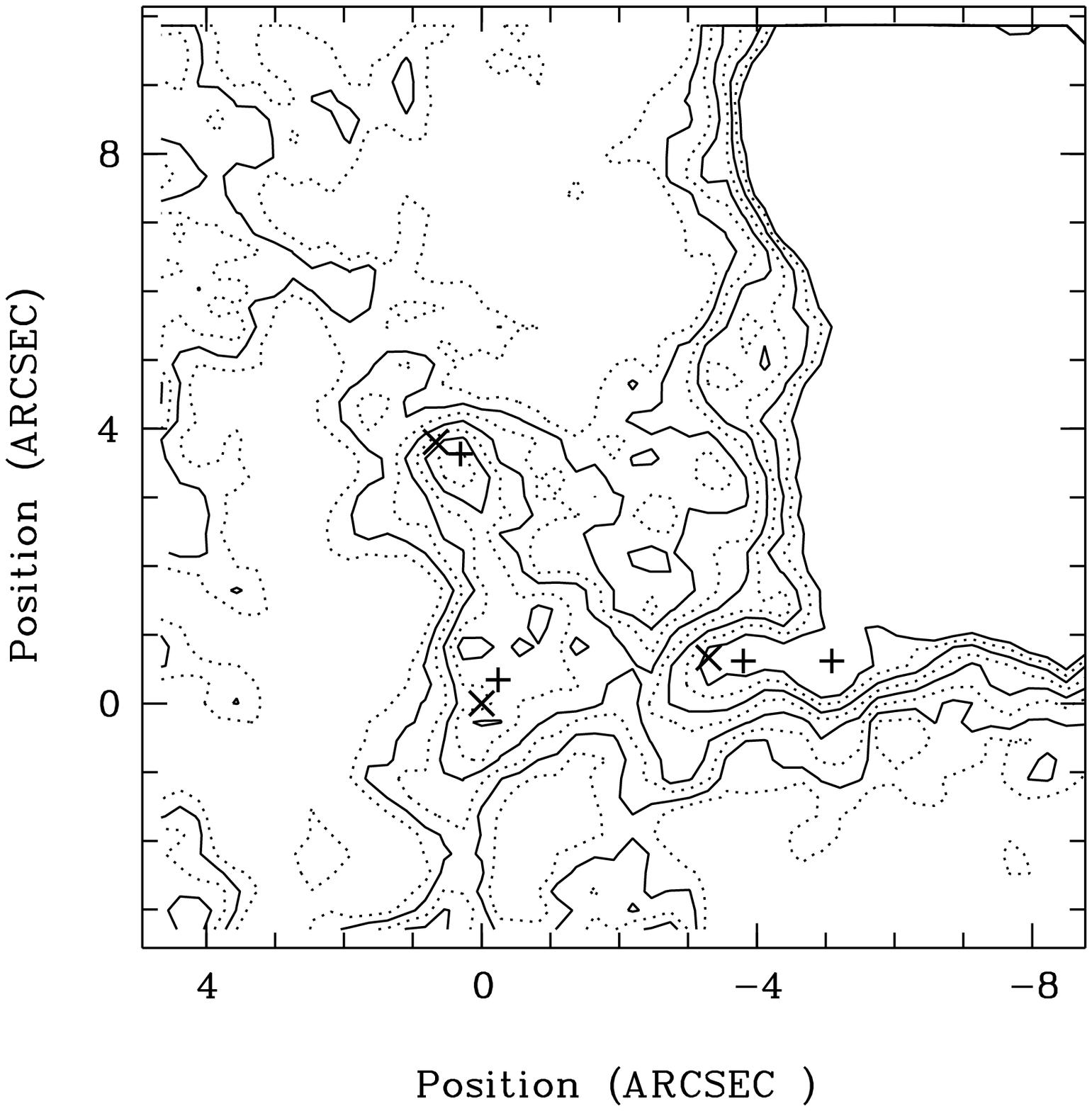}}
\end{picture}
}
\hfill
\parbox[b]{55mm}{
\label{f:656_bta}
\caption{
{\it Left:}
Fragments of the BVRI
images of the \psh\ field taken 
with the BTA during two nights, 
in 1996 and 1997. The images are smoothed 
over the area of $3\times3$ pixels. 
The fragment sizes are 14\arcsec$\times$14\arcsec.  
\psh\ (PSR) and the objects in its vicinity (o2--o5) are marked 
in the R image. 
{\it Right:} The corresponding contour maps.
Crosses (``$\times$'') mark the PSR and o2-o5 objects  
centered at their positions in the \hst\ WFPC2 image.  
Pluses (``+'') mark the positions obtained by the 
{\zz center/gauss} fit of the point-like objects on the 
BTA images (see Sect.~\ref{s:flux} for details).  
The origin of the reference frame
is placed at the WFPC2 pulsar's ``$\times$''-position;
the axes are scaled in arcseconds. }} 
\end{figure*}

\section {Observations and data analysis}

\label{s:observations}

\subsection{BTA observations}

We include in our analysis the data obtained in two
BTA observations -- of 1996, November 11/12 (described
by \cite{Kurt}) and 1997, November 26/27.
Both observations were carried out
with a CCD photometer, with a pixel size of
$0\farcs274 \times 0\farcs274$.
The response curves of the filters used 
are close to those of the
B, V, R$_c$, and I$_c$ filters
of the Johnson-Cousins system (referred to as BVRI hereafter).
Other technical details can be found in \cite{Kurt}.

The 1997 observations were performed with the R and I filters only.
Because of poor seeing and high sky background,
we had to exclude from consideration  seven of ten images taken 
with the I filter and three of five images obtained with the R filter.
Observational conditions for the accepted exposures
are presented in Table~\ref{t:obs}.

We reduced the data using
the ESO {\zz MIDAS} and STScI {\zz IRAF} software.
We summed the new R and I images with the corresponding images taken on 
November  1996.  This decreases  the statistical errors
of flux measurements 
in comparison with the previous measurements based 
on 1996 data alone.  

The instrumental stellar magnitudes 
$m^{\mathrm{instr}}$ 
for each filter (B,V,R,I) are 
\beq
m^{\mathrm{instr}} = -2.5 \log \left(\frac{f}{t_{\mathrm{exp}}}\right) - 
\delta m - \frac{k}{\cos Z}~,
\label{eq:m}
\eeq
where $f$ is the flux in counts (DN) for a given aperture,
$\delta m$ 
is the correction 
for finite aperture derived from the point spread function (PSF) of
bright stars, 
$k$ is the extinction 
factor, for which we used average values of the BTA observatory: 
$k=0.34$, 0.21, 0.15, and 0.10 for the B, V, R, and I filters,
respectively (\cite{Neizv}).
The signal-to-noise ratios $S/N$ and 
the magnitude uncertainties $\Delta m^{\mathrm{instr}}$ were calculated as
\beq
\frac{S}{N} =
f \left[\frac{f}{G} + A\, 
\sigma_f^2
\left(1+\frac{A}{A_{\mathrm{sky}}}\right)\right]^{-1/2}, 
\eeq
\beq
\Delta m^{\mathrm{instr}} =
\frac{2.5}{\ln 10} \frac{N}{S},
\label{eq:SNDm}
\eeq
where $\sigma_f$ is
the standard deviation of flux in counts, 
$A$ is the number of pixels in the source aperture,
$G$ is the gain,
and 
$A_{\mathrm{sky}}$ is the number of pixels 
used for background measurement.
The instrumental magnitudes $b,v,r,i$ were then transformed to the 
Johnson-Cousins magnitudes $B,V,R,I$ using the equations
\beq
\eqalign{
B - b & = 26.16(3) + 0.08(3) \cdot (b - v)~, \cr
V - v & = 26.28(2) - 0.11(3) \cdot (b - v)~, \cr
R - r & = 26.59(2) + 0.02(1) \cdot (v - r)~, \cr
I - i & = 25.67(2) + 0.06(2) \cdot (r - i)~, \cr
        }
\label{eq:p6pheqs}
\eeq
where the transformation coefficients were determined with the aid
of secondary photometric standards described by \cite{Kurt}.
The absolute fluxes 
$F$ 
(in erg cm$^{-2}$ s$^{-1}$ Hz$^{-1}$)
were calculated using the following equation 
\beq
\log F = - 0.4 \left( m + m^0 \right),
\label{eq:fl_mag}
\eeq 
with the zero-points provided by \cite{Fkg}:
\beq
\eqalign{
m_B^0 = 48.490,\;\; m_V^0 = 48.613, \cr
m_R^0 = 48.800,\;\; m_I^0 = 49.058. \cr
        }
\label{eq:mag2fl_bta}
\eeq

\begin{figure*}[t]
\setlength{\unitlength}{1mm}
\begin{picture}(175,90)(0,3)
\put ( 0,11) {\includegraphics[width=77mm,bb=70 303 522 
755,clip]{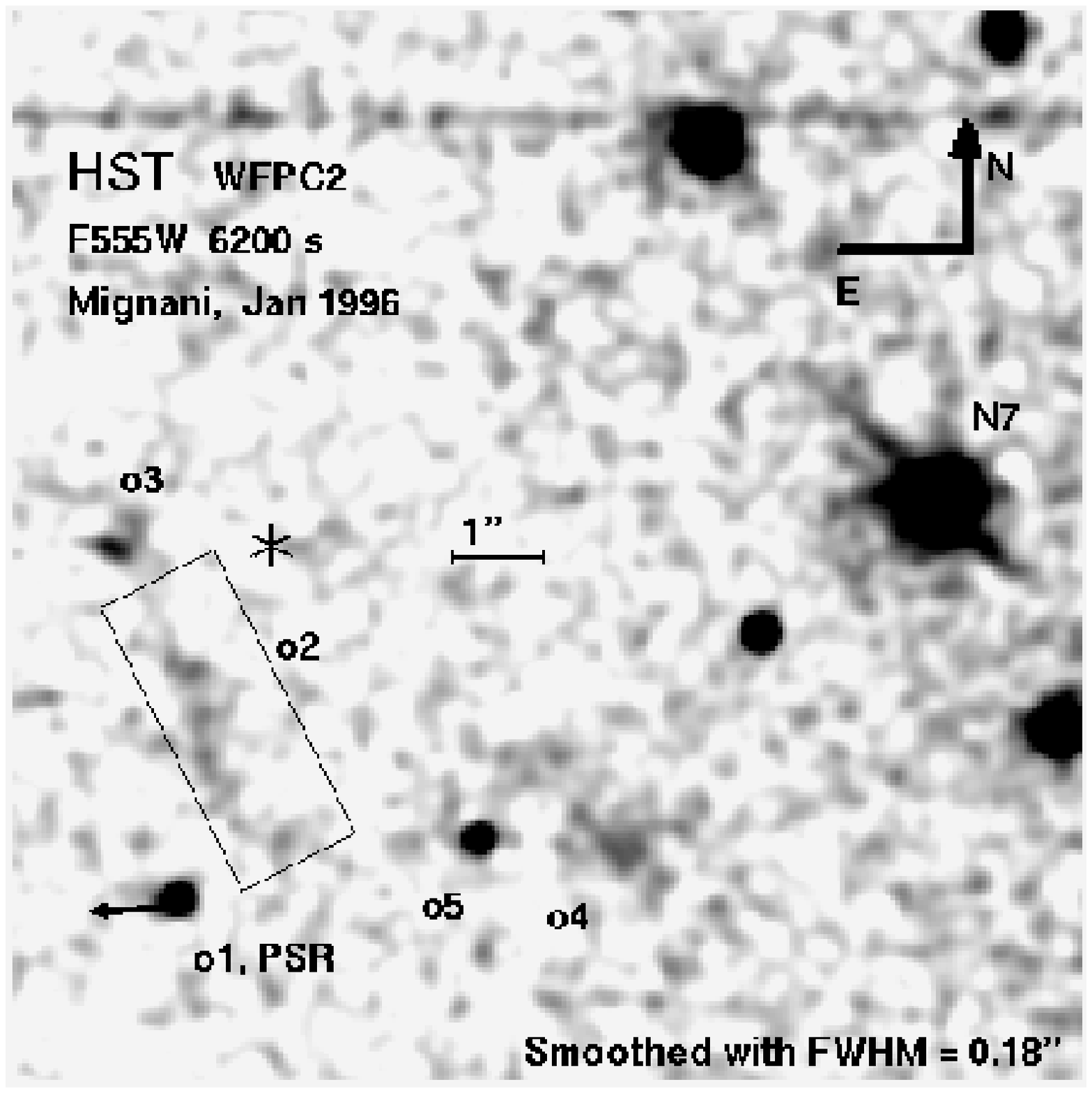}}
\put (80,0 ) {\includegraphics[width=95mm,bb=70 355 500 
785,clip]{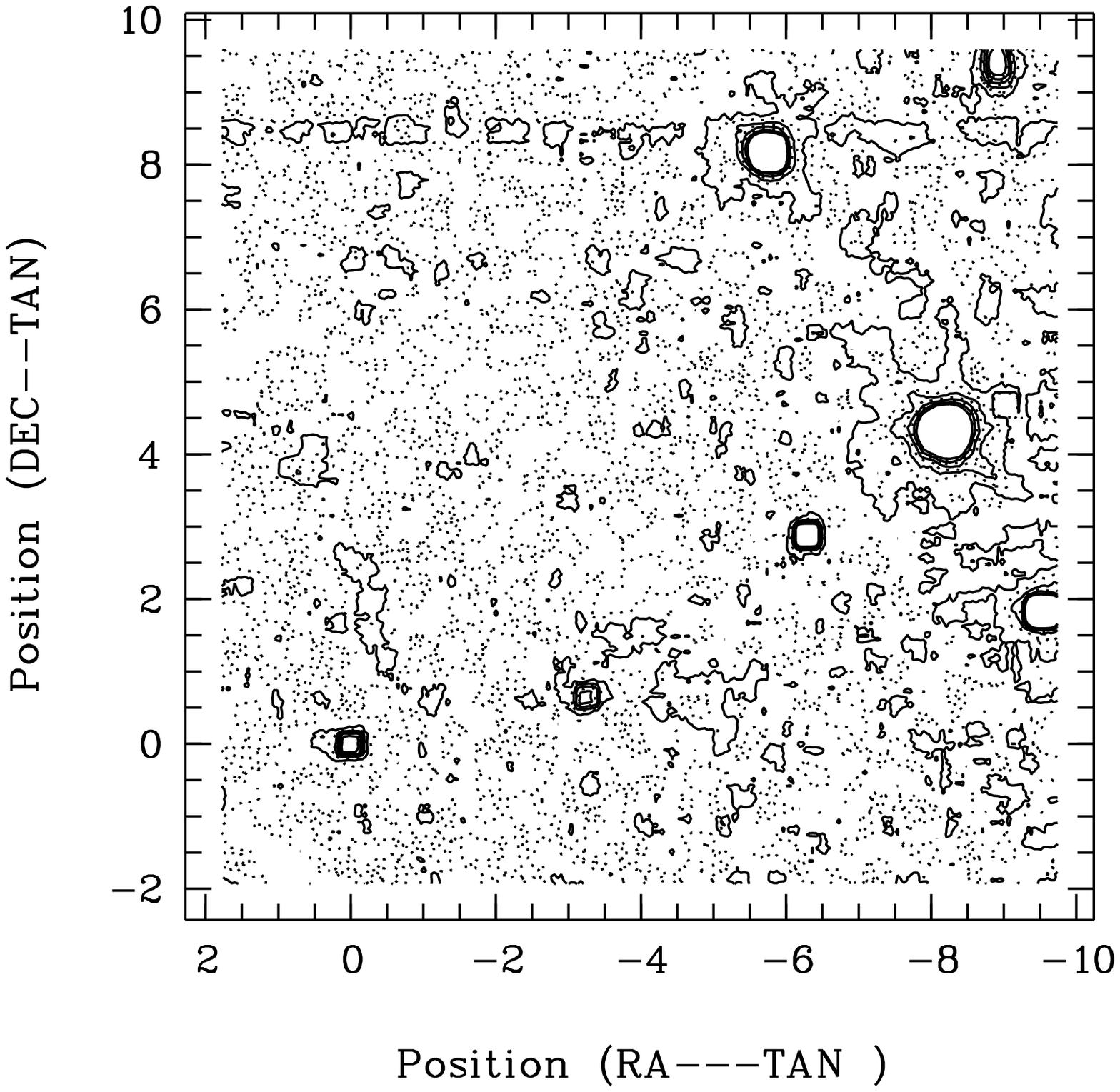}}
\end{picture}
\caption{
WFPC2 image of the pulsar and surrounding field.
The fragment is 11\farcs6$\times$11\farcs6\ in size and   
smoothed with $F\!W\!H\!M = 0$\farcs18.
The arrow in the left panel
shows the direction of the pulsar proper motion 
(\cite{mPM}),
its length corresponds to the distance traveled by the pulsar in 20
years.
The rectangle and the asterisk are discussed in Sect.~\ref{contam}.}
\label{f:656_hstW}
\end{figure*}

\begin{figure*}[p]
\begin{center}
\setlength{\unitlength}{1mm}
\begin{picture}(140,215)(0,7)
\put ( 0,13.5)   {\includegraphics[width=58.3mm,bb=70 303 522 
755,clip]{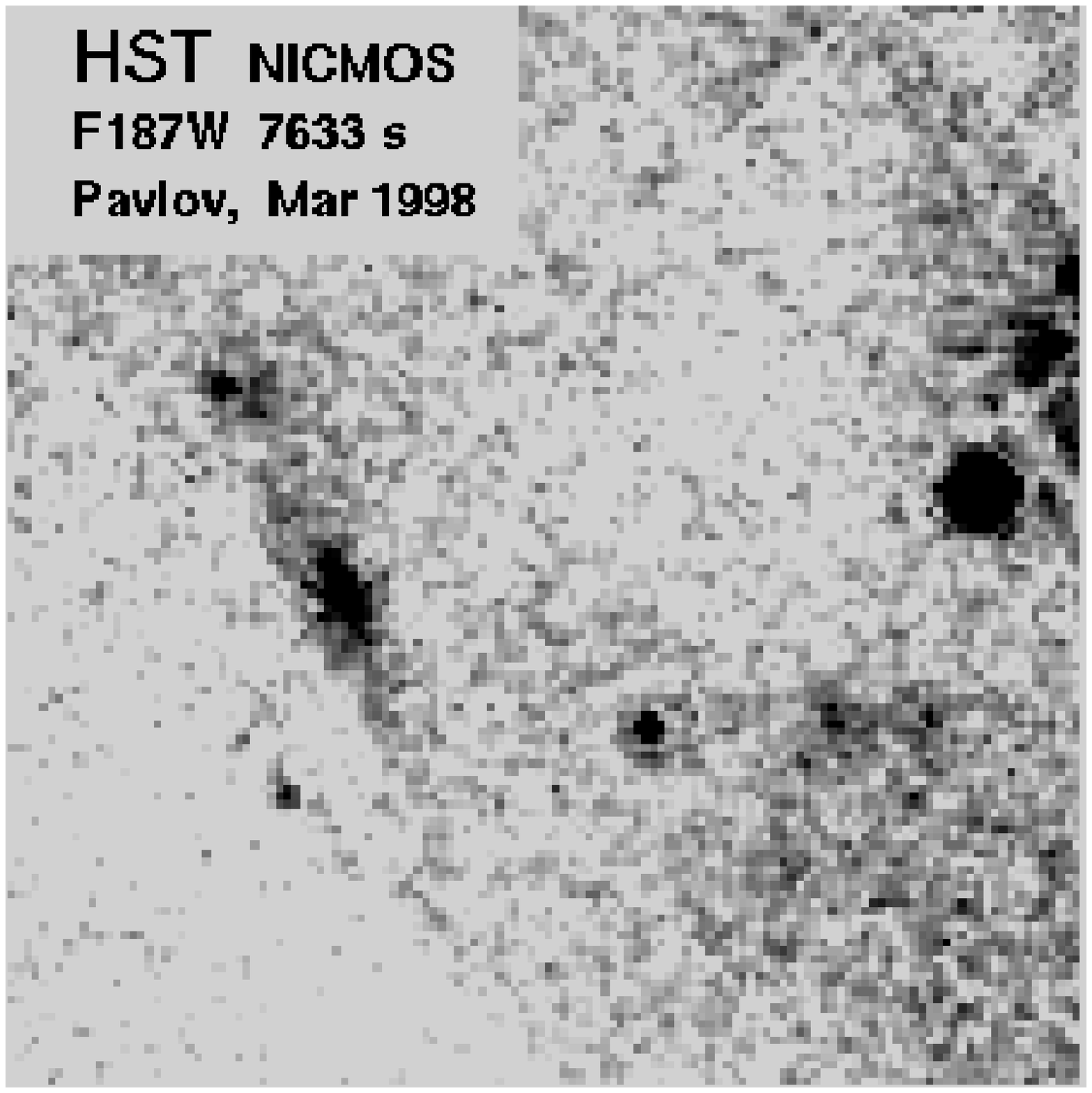}}
\put ( 0,85.5)   {\includegraphics[width=58.3mm,bb=70 303 522 
755,clip]{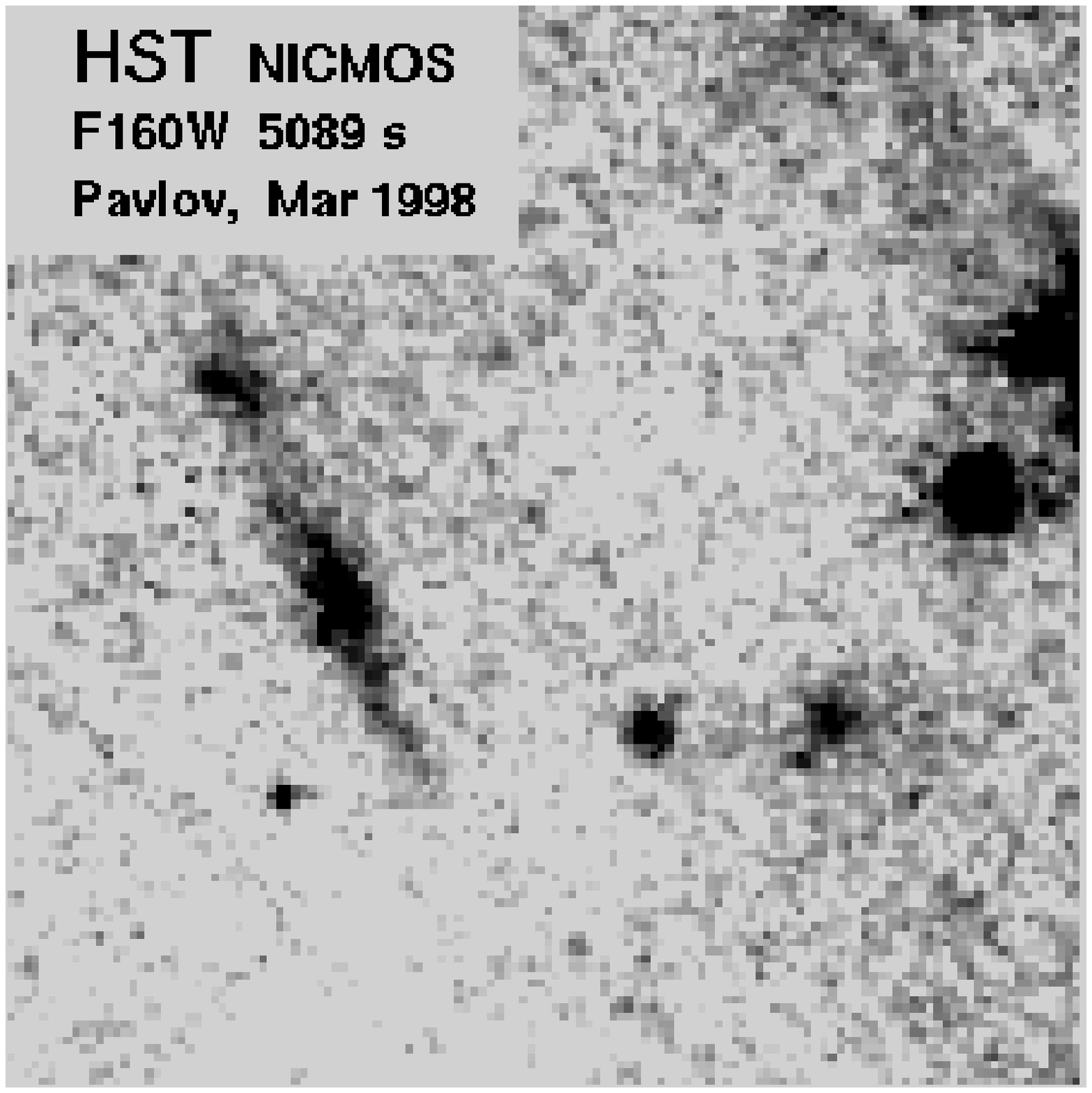}}
\put ( 0,157.5)   {\includegraphics[width=58.3mm,bb=70 303 522 
755,clip]{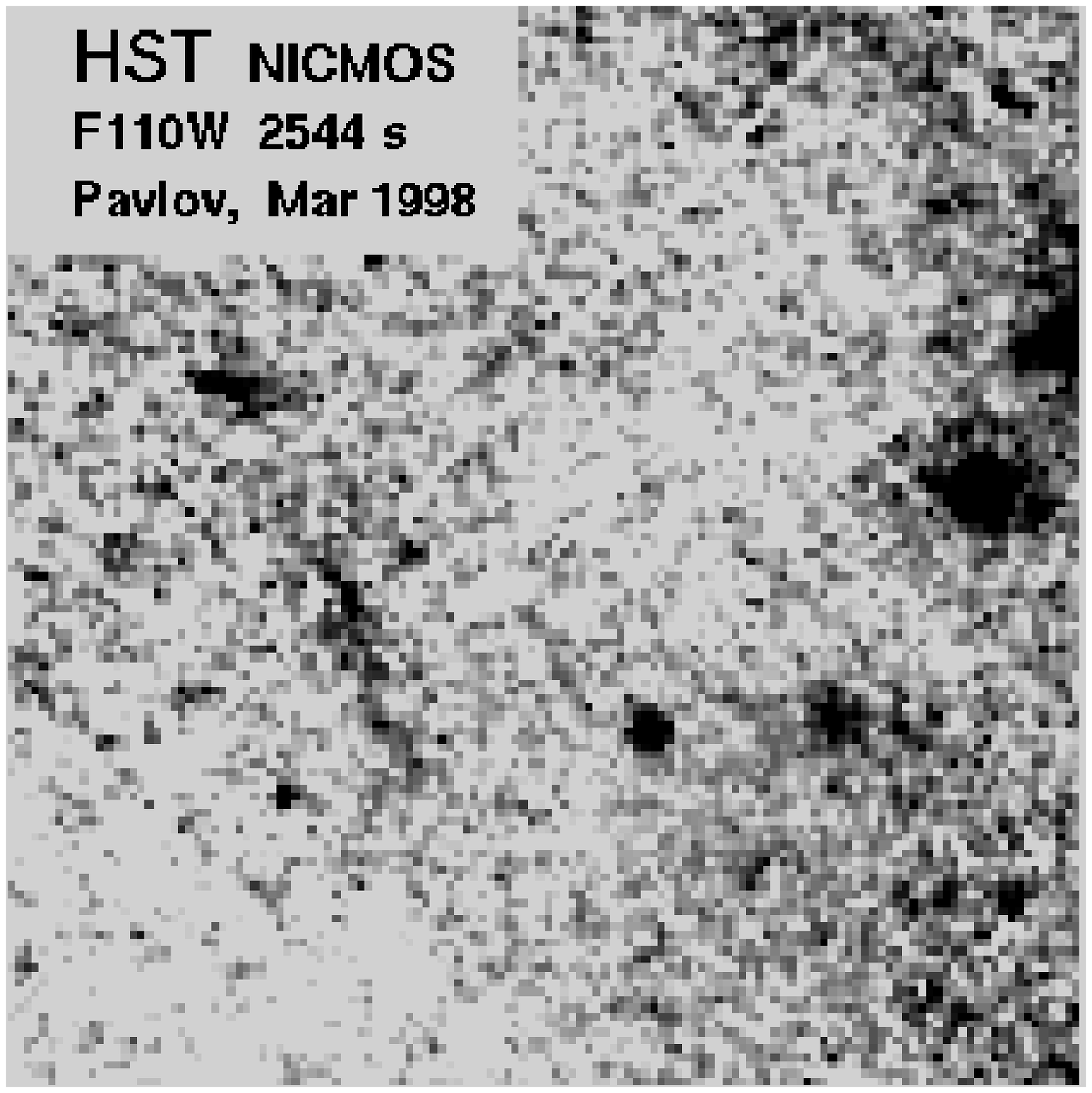}}
\put (65,  2.9) {\includegraphics[width=72.81mm,height=72.45mm,bb=80 325 535 
775,clip]{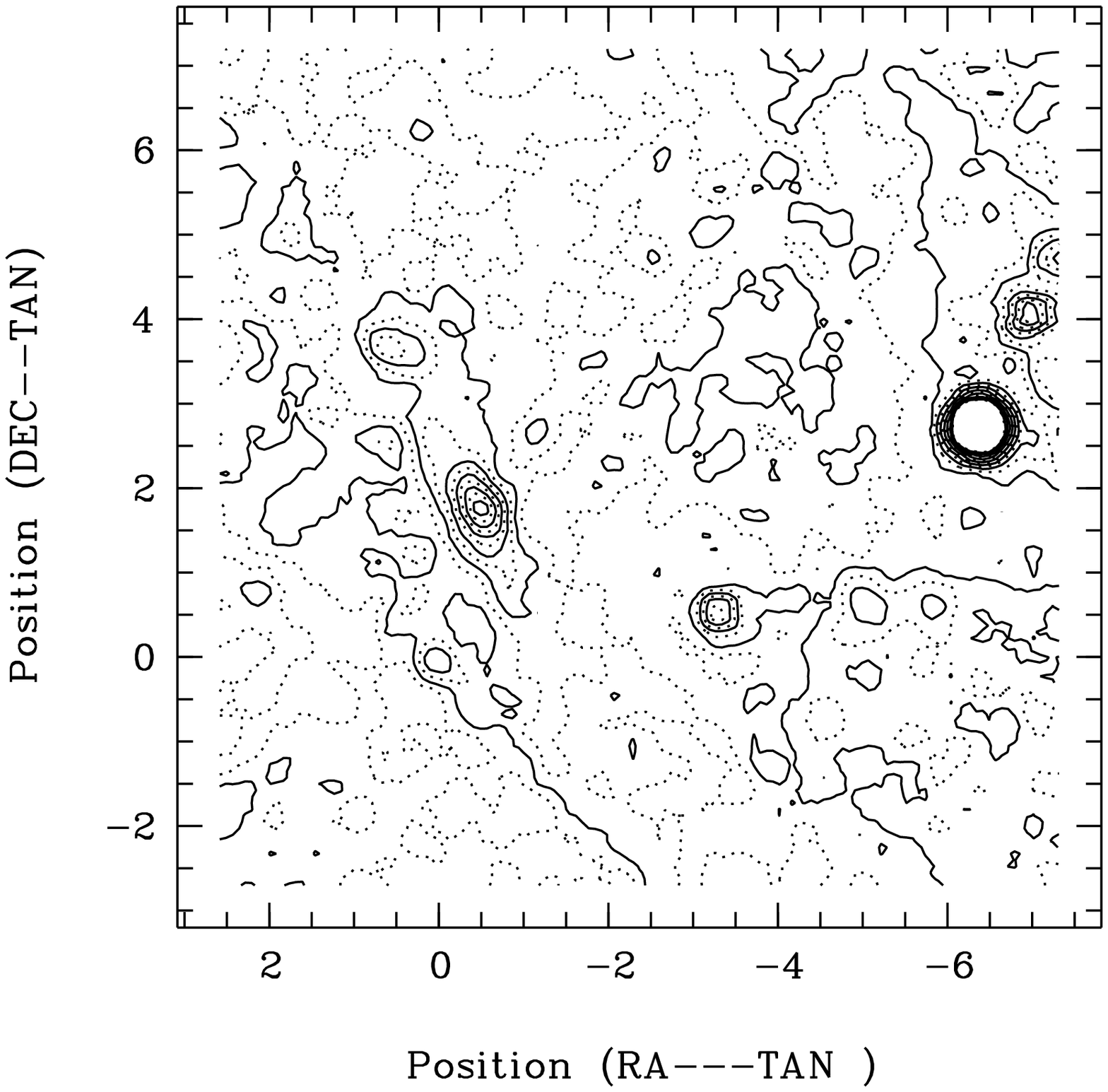}}
\put (65, 74.9) {\includegraphics[width=72.81mm,height=72.45mm,bb=80 325 535 
775,clip]{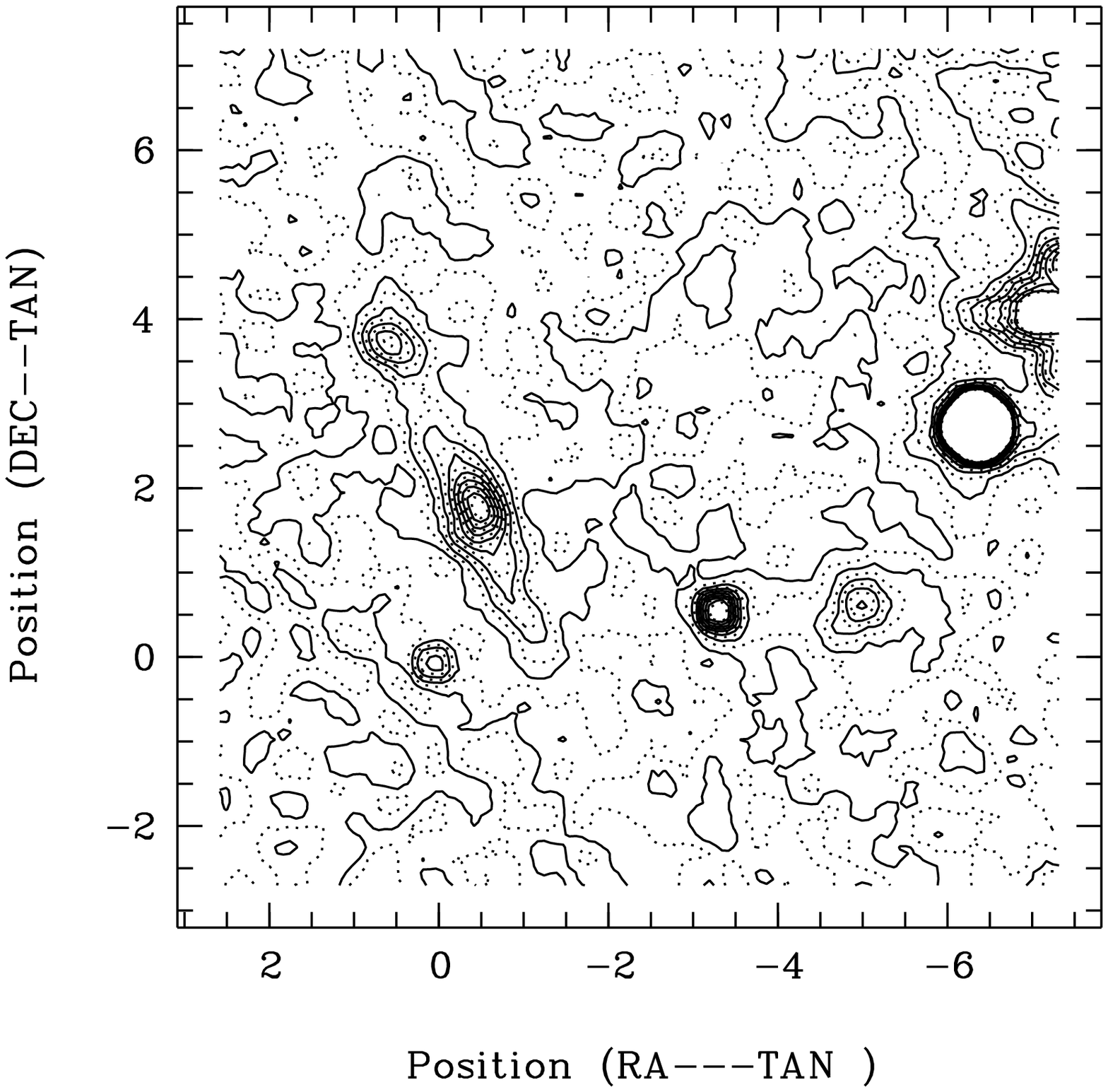}}
\put (65,146.9) {\includegraphics[width=72.81mm,height=72.45mm,bb=80 325 535 
775,clip]{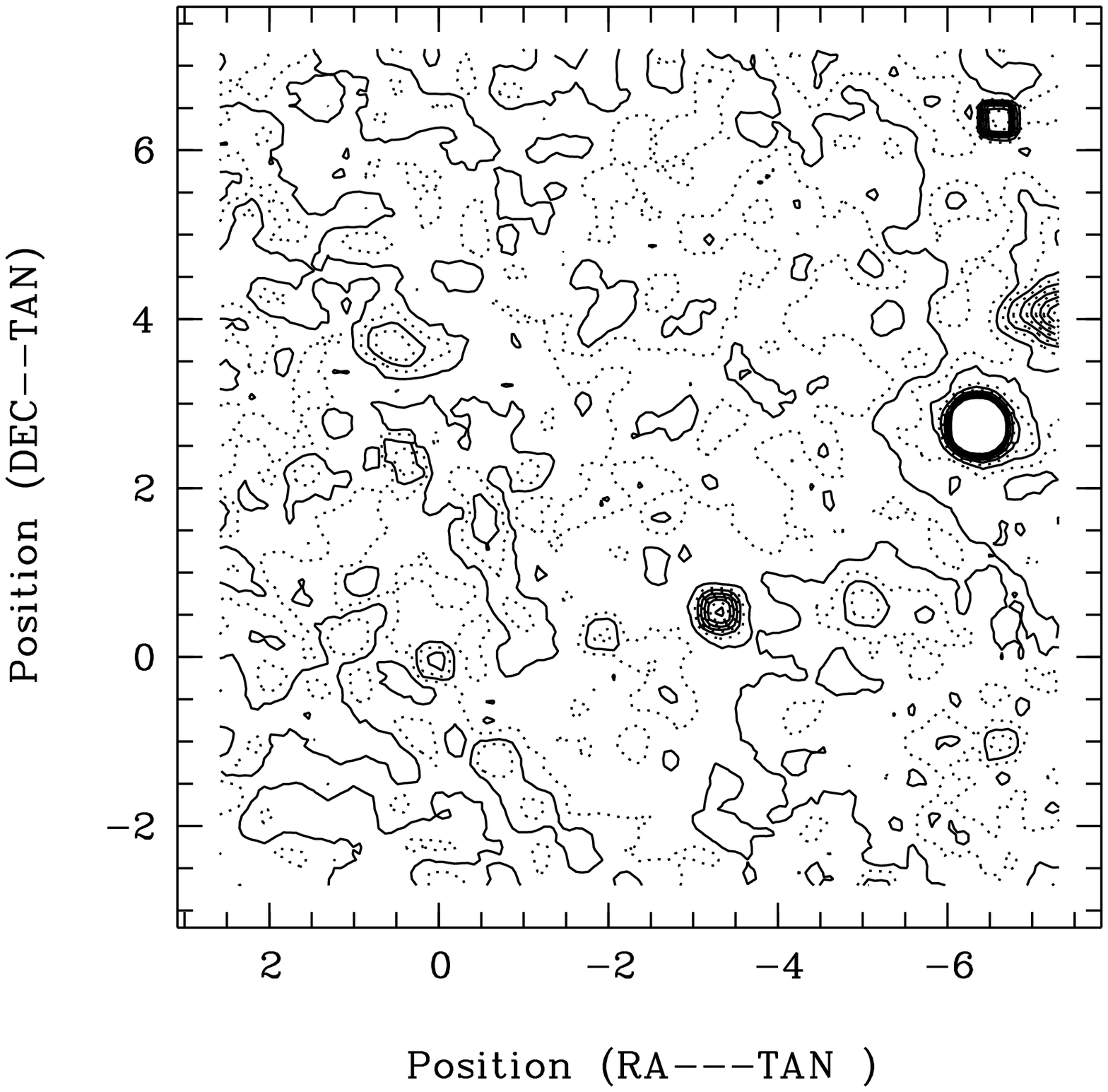}}
\end{picture}
\end{center}
\caption{
The same as in Figs.~\ref{f:656_bta}, \ref{f:656_hstW} 
but for the \hst/NICMOS bands. 
The fragments represent the region  10\arcsec$\times$10\arcsec\ in size
of the non-smoothed images.  
} 
\label{f:656_hstN}
\end{figure*}

\subsection{HST data}

We also studied the \psh\ field using the 
\hst\ data obtained by \cite{Mign} 
in the WFPC2 F555W band   and by \cite{Harlow} 
in the NICMOS F110W, F160W and F187W bands. 
The standard reduction scheme described in the \hst\ 
Data Handbook  
was applied to the WFPC2 images processed by pipeline. 
The data were processed with the STScI {\zz IRAF} software, 
summed with the {\zz IRAF} task {\zz crrej},
and a $\sim$~116\arcsec~$\times$~116\arcsec\ mosaic
image
containing three WFC and one PC field,
corrected for the \hst\ geometric distortion, was constructed.
We applied the photometric formulae similar to 
Eqs.~(\ref{eq:m}) and (\ref{eq:SNDm}), with allowance for
the WFPC2 cooldown and decontamination corrections.
The correction for the finite aperture was performed with the aid of
the PSF image {\zz gb71613du} from the WFPC2 PSF Library. 

For the primary reduction of the NICMOS data, we used the standard 
procedure
{\zz calnica}\footnote{We made use of the calibration files
{\zz i7l12297n}, {\zz h421621an}, {\zz h4214599n},  
{\zz i7l1653gn},  
{\zz h2413242n}, {\zz h241323pn}, {\zz h241323qn}, {\zz icb15080n}, {\zz 
icb1507pn}, 
and {\zz icb1507qn}, and the dark calibration file created by Brian Monroe
(NICMOS Helpdesk at STScI).}.
For rejection of anomalously ``cold'' pixels, we applied the
{\zz IRAF} task {\zz cosmicrays} to inverted images.
To remove the pedestal effect\footnote{See
\href{http://www.stsci.edu/instruments/nicmos/anom_pedestal.html}
{http://www.stsci.edu/instruments/nicmos/anom\_pedestal.html}}
and to subtract the background,
the {\zz IRAF} task {\zz pedsky}\footnote{See
STAN (Space Telescope Analysis Newsletter) No. 18. \newline
\href{http://www.stsci.edu/instruments/nicmos/NICMOS_stans/stan_018.ascii}
{www.stsci.edu/instruments/nicmos/NICMOS\_stans/stan\_018.asci}}
was implemented,   
with the same reference flatfield.
It significantly suppressed the pedestal level,
but increased the background level 
near the edges of the images,
which is not critical for our purposes since the objects of interest 
are near the image centers.
To combine the individual dither images
and get rid of cosmic rays, we 
made use of the package {\zz ditherII\/} by \cite{dither}
and followed the recipe
described in the example supplied with the package 
release\footnote{\href{http://www.stsci.edu/~fruchter/dither/ditherII.ps}
{http://www.stsci.edu/$\sim$fruchter/dither/ditherII.ps}}.
The shifts of single images were calculated
making use of three bright stars in the field.
The PSF images were generated by the TinyTim 
code\footnote{\href{http://www.stsci.edu/software/tinytim}
{http://www.stsci.edu/software/tinytim}}.
Aperture sizes were chosen to maximize the signal-to-noise ratio.

To estimate the flux errors for the sources observed with the
\hst, we neglected possible systematic uncertainties
which are smaller than statistical errors for faint objects,
including the pulsar. Conversion of the fluxes to the 
Vega system magnitudes for the \hst\ bands 
were performed using  Eq.~(\ref{eq:fl_mag}) and 
the following  zero-points, calculated by the pipeline:
\beq
\eqalign{
m_{555}^0 = 48.610,\;\; m_{110}^0 = 49.305, \cr
m_{160}^0 = 49.885,\;\; m_{187}^0 = 50.148. \cr
        }
\label{eq:mag2fl_hst}
\eeq

\subsection{Astrometry}

\label{CoordTransf}

For astrometrical referencing of the BTA images, 
we used the mosaic WFPC2/F555W image
which covers a fraction of the BTA field. 
The coordinates of five relatively bright stars 
in this overlapping region
were determined 
in the WFPC2 and  BTA images. 
Then, a six-parameter
coordinate transformation was constructed.
The uncertainties of the transformation are equal to
$\approx 0\farcs07$ for the R,I images and 
0\farcs15 ($\sim$~0.5 pixel) 
for the B,V images. 
This allows us to relate the BTA and \hst\ images  
with accuracy better than the BTA pixel size.

\section{Multicolor photometry of \psh\ and objects in its neighborhood}

\label{s:photometry}

\subsection{Morphology of \psh\ field in different 
bands}

14\arcsec$\times$14\arcsec\ fragments
of the smoothed BTA BVRI images\footnote{These and other
images are available in {\zz FITS} format at
\href{http://www.ioffe.rssi.ru/astro/NSG/obs/0656-phot.html}
{http://www.ioffe.rssi.ru/astro/NSG/obs/0656-phot.html}}
containing the pulsar are presented in Fig.~\ref{f:656_bta}.
In Figs.~\ref{f:656_hstW} and 
\ref{f:656_hstN} we show 
a smoothed 11\farcs6$\times$11\farcs6 fragment  
of the \hst/WFPC2 image 
and non-smoothed
10\arcsec$\times$10\arcsec\ fragments 
of the \hst/NICMOS images, respectively. 
All the images are oriented as indicated
in Fig.~\ref{f:656_hstW}.
The contour plots of 
the images in Figs.~\ref{f:656_bta}, \ref{f:656_hstW} and \ref{f:656_hstN}
are smoothed over the areas 
of 3$\times$3, 7$\times$7, and 5$\times$5 pixels for 
the BTA, WFPC2, and NICMOS images, respectively.
The isophotes correspond to the following 
levels $l_n$ (in counts) above the background: 
\beq
l_n = S + n k \sigma,
\eeq
where $S$ is the mean sky value near the pulsar,
$n = 0,1,\ldots$ $N$;
$\sigma$ is the  sky standard deviation related to one pixel;
$k$ is the scaling factor. We used
$k$ = 1.25, 2 and 3, $N$ = 10, 10 and 18, for the
BTA, WFPC2 and NICMOS contour plots, respectively.
All the images confirm the detection by \cite{Kurt}
of five faint objects, o1--o5
(marked in Figs.~\ref{f:656_bta} and~\ref{f:656_hstW}),
within $\sim 5$\arcsec\ around the pulsar position.

\begin{table*}[t]
\caption{Flux measurements of the pulsar's optical counterpart. }
\label{t:656_phot}
\vbox{
\begin{tabular}{c|cccclcc}
\hline\hline
Band         & $S/N$ & $r^a$  & Source$^b$ & $k/\cos(Z)^c$ & 
$\delta_{\rm fin}^d$ & Mag$^e$     & Flux \widerul  \\
             &       & pix     & DN s$^{-1}$ & mag            &                  &           & $\mu$Jy \\
\hline
B$^+$        & 10    &  3      & 0.839(80)   & 0.41           & 2.44             & 24.98(13) & 0.409(36) \widerul \\
V$^+$        & 9.0   &  3      & 1.26(14)    & 0.25           & 2.07             & 24.97(12) & 0.369(39) \\
R$^+$        & 16    &  3      & 2.77(17)    & 0.18           & 2.15             & 24.48(7)  & 0.488(32) \\
I$^+$        & 10    &  3      & 2.26(22)    & 0.12           & 2.05             & 23.87(11) & 0.675(65) \\
\hline
B$^\times$   & 8.1   &  3      & 0.661(81)   & 0.41           & 2.44             & 25.25(13) & 0.319(36) \\
V$^\times$   & 9.2   &  3      & 1.25(14)    & 0.25           & 2.07             & 24.97(12) & 0.369(39) \\
R$^\times$   & 11    &  3      & 2.24(19)    & 0.18           & 2.15             & 24.71(9)  & 0.394(31) \\
I$^\times$   & 9.5   &  3      & 2.02(21)    & 0.12           & 2.05             & 23.99(11) & 0.604(58) \\
\hline
F555W        & 23    &  4      & 0.0853(38)  & -              & 1.12             & 24.90(7)  & 0.393(24) \\
F110W        & 11    &  2      & 0.119(11)   & -              & 1.55             & 24.38(10) & 0.336(30) \\
F160W        & 14    &  3      & 0.185(13)   & -              & 1.50             & 23.22(8)  & 0.575(41) \\
F187W        & 8.0   &  3      & 0.127(16)   & -              & 1.65             & 22.62(13) & 0.779(97) \\
\hline
\end{tabular}
}{
\renewcommand{\arraystretch}{0.8}
\begin{tabular}{cl}
& \\
$^+$      & Aperture centered at the position determined by the {\zz center/gauss} fit in the BTA images \\
$^\times$ & Aperture centered 
at the positions determined from the WFPC2 image,
(see `+' and `$\times$' marks in Fig.~\ref{f:656_bta})\\
$^a$      & Aperture radius in pixels \\
$^b$      & Source count rate within the aperture   \\     
$^c$ & Correction for atmospheric extinction \\
$^d$ & 
Factor to correct for the finite aperture \\
$^e$ & Johnson-Cousins and Vega magnitudes,
for the BTA and \hst\ images, respectively \\
\end{tabular}
}
\end{table*}

Because of worse seeing during the BTA observations, we cannot
determine from the BTA images whether the faint 
objects are point-like or extended, except for o2 which is  
clearly extended and looks as associated with the pulsar. 
With the aid of the WFPC2 and NICMOS images,
we infer that the pulsar counterpart o1
($\alpha_{2000} = 6^h 59^m 48\fs19$,
$\delta_{2000}=14^\circ 14\arcmin 21\farcs3$,
as determined from the NICMOS images\footnote{This and other
absolute positions in the \hst\ images are given as directly measured
with the {\zz IRAF imcntr} task, without correcting for possible \hst\
pointing errors.})
and the stellar object o5
($\alpha_{2000} = 6^h 59^m 47$\fs96, $\delta_{2000}$ =
14$^\circ$14\arcmin 21\farcs9;
$\simeq 3\arcsec$ west of the pulsar) have 
point-like source profiles,   
whereas o2, o3, and o4 are 
extended. 

The extended sources are  most 
clearly resolved in the NICMOS/F160W band. 
The object o2 shows a compact bright nucleus 
at $\alpha_{2000} = 6^h 59^m 48$\fs16, 
$\delta_{2000}$ = 14$^\circ$14\arcmin 23\farcs2
(1\farcs85 north of the pulsar). 
The nucleus is surrounded by a faint elongated nebula which 
occupies an area of $\sim 3\arcsec \times 0\farcs7$.
The minimum distance between the nebula's southern edge 
and the pulsar, $\simeq $~1\farcs1 in the F160W band, 
is smaller than the BTA seeing. This explains 
naturally the apparent association of the pulsar and 
o2 in the R and I bands. In the F555W band, 
whose response curve is very close to
that of the Johnson's V filter, o2 looks like 
a faint elongated clump without a pronounced nucleus. 
Also, o2 is barely seen 
as a very faint extended source in the FOC/F130LP image 
obtained by \cite{PSC}.   

The extended object o3 
($\alpha_{2000} = 6^h 59^m 48$\fs23, 
$\delta_{2000}$ = 14$^\circ$14\arcmin 25\farcs1) 
is  $\simeq$~4\arcsec\ north of the pulsar.
It has an irregular structure with a bright knot at its eastern edge. 
Being at only $\sim$~1\arcsec\ from the northern edge of o2,
the object o3 appears to be associated 
with o2 in the F160W  image. 
The structure of the object o4 ($\alpha_{2000} = 6^h 59^m 47$\fs84, 
$\delta_{2000}$ = 14$^\circ$14\arcmin 22\farcs0;
$\sim$~5\arcsec\ west of the pulsar)
varies strongly from band to band.

\subsection{Consistency of the BTA and HST data}
 
To check consistency of our data reduction and photometry
of the objects observed with the different telescopes and detectors,
we measured the magnitudes of a relatively bright star 
(marked N7 in Fig.~\ref{f:656_hstW},
following the notation by \cite{Kurt})
seen in all the BTA and \hst\ images. 
The raw broad-band spectrum of this star 
(Fig.~\ref{f:o5N7} and Table ~\ref{t:obj})
looks rather smooth and shows no  
jumps between the BTA and \hst\ bands. 
Moreover, the fluxes in the almost equivalent 
V and F555W bands are in good agreement with each other.
This indicates that
there are no major inconsistencies
between the reduced BTA and \hst\ data. 
The colors and magnitudes of N7
correspond to an M5V star (\cite{Bes90})
at a distance of 440 pc. 

\subsection {Flux measurements} 

\begin{table*}[t]
\caption{Fluxes (upper values, in $\mu$Jy)
and Johnson-Cousins (BVRI) or Vega 
magnitudes (lower values) of the objects in the vicinity of \psh.}
\label{t:obj}
\begin{tabular}{c|cccccccc}
\hline\hline
Object  & B                                 & V                                 & F555W                               & R                                 & I                                 & F110W                              & F160W                              & F187W \widerul \\
\hline
o2      & \twolines{0.122(34)}{26\fm29(30)} & \twolines{0.187(38)}{25\fm71(22)} & \twolines{0.36(12)}{25\fm00(36)}    & \twolines{0.211(32)}{25\fm39(16)} & \twolines{0.526(59)}{24\fm14(12)} & \twolines{1.14(14)}{23\fm06(13)}   & \twolines{3.58(15)}{21\fm23(4)}    & \twolines{4.93(27)}{20\fm62(6)}         \\
o3      & \twolines{0.205(36)}{25\fm73(19)} & \twolines{0.282(39)}{25\fm26(15)} & \twolines{0.35(11)}{25\fm03(34)}    & \twolines{0.353(34)}{24\fm83(10)} & \twolines{0.512(62)}{24\fm17(13)} & \twolines{0.77(16)}{23\fm49(23)}   & \twolines{1.66(15)}{22\fm07(10)}   & \twolines{2.40(27)}{21\fm04(12)}        \\
o4      & \twolines{0.217(38)}{25\fm67(19)} & \twolines{0.136(34)}{26\fm05(27)} & \twolines{0.42(12)}{24\fm83(31)}    & \twolines{0.398(32)}{24\fm70(9)}  & \twolines{1.099(69)}{23\fm34(7)}  & \twolines{1.00(13)}{23\fm19(14)}   & \twolines{1.64(12)}{22\fm08(8)}   & \twolines{1.44(25)}{21\fm95(19)}        \\
\hline
o5      & -                                 & -                                 & \twolines{0.268(20)}{25\fm32(8)}    & \twolines{0.520(41)}{24\fm41(9)}  & \twolines{0.774(75)}{23\fm72(11)} & \twolines{0.991(33)}{23\fm21(4)}  & \twolines{1.488(37)}{22\fm18(3)}  & \twolines{1.374(78)}{22\fm01(6)}        \\
N7      & \twolines{6.37(34)}{22\fm00(6)}   & \twolines{26.7(12)}{20\fm32(5)}   & \twolines{23.364(40)}{20\fm469(2)}  & \twolines{64.9(12)}{24\fm17(2)}   & \twolines{192.8(52)}{17\fm73(3)}  & \twolines{313.99(25)}{16\fm954(1)} & \twolines{439.96(25)}{16\fm008(1)} & \twolines{386.65(30)}{15\fm884(1)}        \\
\hline
\hline
\end{tabular}
\end{table*}


\label{s:flux}

The objects of interest
are faint and strongly blurred in the BTA images,
especially in the R and I filters. 
This leads to additional uncertainties of their  
center positions and fluxes.
To evaluate the uncertainties, we determined the positions
(and the corresponding fluxes) of the unresolved objects
using two approaches.
First, we applied the
{\zz center/gauss} task of the {\zz MIDAS} package
to our BTA images. 
These positions are marked by the `+' signs at the contour 
plots in Fig.~\ref{f:656_bta}.
(The center of o5 cannot be determined by this method  
in the B and V images because o5 is 
hidden in the background.)
Second, we measured the positions of o1 (PSR), o5 and o3
in the WFPC2 image and transformed them to the BTA positions
as described in the Sect.~\ref{CoordTransf}.
These positions are marked by the `$\times$' signs
(o4 cannot be centered in this way because it is resolved
in the WFPC2 image.) We see from Fig.~\ref{f:656_bta} that
the `+' positions of the point sources 
(o1 and o5) are systematically closer,
by 1--2 BTA pixels, to the
nearest extended objects (o2 and o4, respectively) than the
`$\times$' positions. 
This means that the fluxes of the faint, point-like sources 
may be contaminated  
by the nearby extended objects,
and the Gaussian fits of their profiles  
may not be reliable.
   
We measured the
stellar magnitudes and fluxes of point-like objects
for a range of aperture radii $r$.
For each of these objects in the BTA images, we chose 
$r=3$ pixels as an optimal value
and measured the background over the annulus
with inner and outer radii $r_{\rm in}=8$ and $r_{\rm out}=11$ pixels, 
centered at the source. To measure the fluxes
in the \hst\ images, we used the task {\zz phot} from 
the {\zz IRAF} package {\zz apphot}.  
For the NICMOS images (pixel size is 0\farcs075), 
we chose $r=3$ pixels
(except for the pulsar in F110W -- see Table~\ref{t:656_phot}), and
$r_{\rm in}=6$, $r_{\rm out}=9$ pixels
($r_{\rm in}=12$, $r_{\rm out}=15$ pixels for N7). We used
$r=4$, $r_{\rm in}=15$, $r_{\rm out}=20$ pixels
for the WFPC2 image (pixel size is 0\farcs045).
The results  are presented in Tables~\ref{t:656_phot} 
and \ref{t:obj}.

\begin{figure}
\includegraphics[width=85mm,clip]{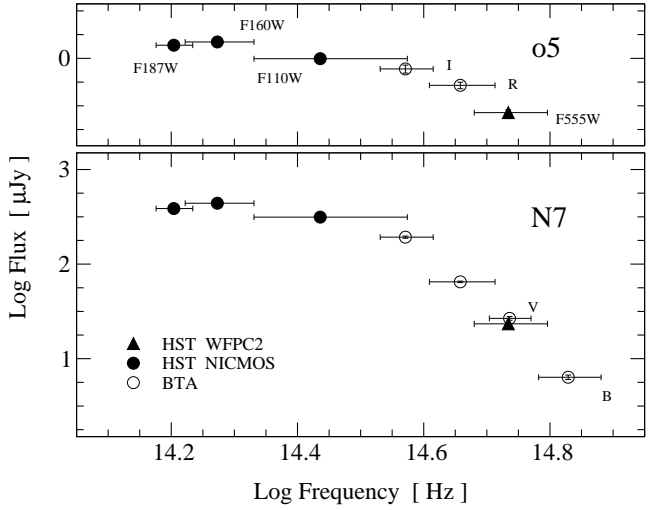}
\caption{Broad-band spectra of the stellar objects o5 and N7.}
\label{f:o5N7}
\end{figure}

We see from Table~\ref{t:656_phot} that
the B, R and I fluxes of the pulsar, measured 
with the apertures centered on the `$\times$' positions
are systematically 
lower, by $\sim$~10--30\%, than those obtained
with the `+' positions. 
In the V band both fluxes coincide with each other and 
are consistent  with the F555W flux, within the errors.
Similar differences between the `$\times$' and `+' fluxes
are found for o3 if it is considered as a point-like source.
Furthermore, the `$\times$' fluxes yield a smooth  
broad-band spectrum of o5   
(see Table~\ref{t:obj} and Fig.~\ref{f:o5N7})
while its `+' flux in the I band
stands out of the smooth spectrum by $\sim 30\%$,
an excess hardly plausible for an ordinary object.
Based on these measurements,
we believe that the `$\times$' fluxes 
of the point-like sources 
are more realistic than the `+' fluxes,
and even the `$\times$' fluxes are likely overestimated
because of contamination from nearby extended sources.

\label{656_shifts_text}
   
We also estimated integral
fluxes of the extended sources in different bands.        
For consistency, the fluxes were measured with the same
circular aperture for all the objects and bands; its
diameter, 1\farcs64, was chosen 
in accordance with the BTA seeing conditions 
in the I band. To measure the fluxes of o2, we 
centered the aperture at
the center of its nucleus in the F160W image
(this position  was transformed into 
the BTA and WFPC2 images using the pulsar and o5 
as reference objects).
Although this aperture does not include the whole o2 nebula, 
the contribution of the outer parts of the nebula 
to the integral flux is only a few percents, 
smaller than the errors of our measurements.   
To estimate the BTA BVRI fluxes, we corrected the 
measured fluxes for the finite aperture 
as if o2 were a point-like object.
Such corrections are negligible for the \hst\ images.
The same approach was used for o3 and o4.  
For these objects, 
the contributions into the measured integral fluxes
from the nearby objects o2 and o5 are not significant.
     
\begin{figure}[t]
\includegraphics[width=85mm,clip]{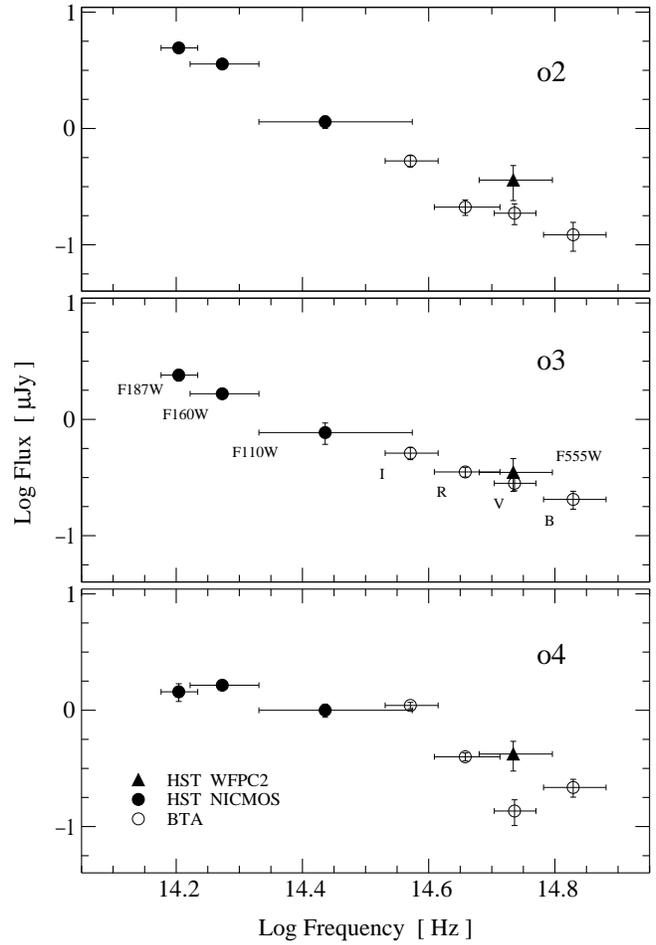}
\caption{Broad-band spectra of the extended objects o2--o4.}
\label{f:o234}
\end{figure}

The magnitudes and the raw broad-band spectra of the extended objects 
are presented in Table~\ref{t:obj} and Fig.~\ref{f:o234}. 
For o2 and o3, the spectra are similar and  
can be crudely approximated 
by a power law, $F_{\nu} \propto \nu^{-\alpha}$, 
with $\alpha \approx 2.4$ and $\approx 1.6$, respectively. 
The spectrum of o4 appears to be more complicated. 
However, the estimated fluxes
of this object are less definite because of
its clumpy morphology and rather low  $S/N$ ($\sim$~1.5--2),
particularly in  the B, V and F555W bands.

\subsection{Contamination of the pulsar flux 
by the object o2 in the R and I bands}

\label{contam}

We see from Tables~\ref{t:656_phot} and \ref{t:obj}
that the `$\times$' fluxes of the pulsar in the R and I bands are comparable
with the corresponding fluxes of o2
whereas the `+' fluxes are higher in both filters. 
The reason is that the `+' positions are shifted towards o2 by 
$\sim$~0\farcs4, so that the `+' fluxes are stronger contaminated.
Therefore, we consider the `$\times$' fluxes more reliable.

If o2 had a point-like profile,
its contribution in the pulsar's aperture would be negligible,
about 3\% in the I band and even less than that in the R band. 
If o2 had the same surface brightness at the pulsar position as 
the brightness measured over the rectangular aperture of 4.6 arcsec$^2$ 
(see Fig.~\ref{f:656_hstW}),
its contribution to the flux measured in the pulsar's aperture
would be 84$\pm$7\% and 96$\pm$6\%
in the R and I bands, respectively.
Since o2 is not a point-like object, and 
it is spatially separated from the pulsar,
the obtained estimates can be considered only 
as lower and upper limits on the contamination.     

An intermediate, more realistic estimate
can be obtained based on the fact that
the brightness distribution of o2 in the F160W image
is almost symmetrical with respect to its major and minor axes.  
We assumed the same source symmetry in the I and R bands, 
and measured the I and R 
fluxes in  the aperture with the radius of 
0\farcs82 (3 pixels)
centered at the point marked by the `{\large $\ast$}' sign
in Fig.~\ref{f:656_hstW}, correcting the fluxes for the finite
PSF as if there were a point source. 
No excess over the background level was detected 
at this position in the R image. 
In the I image we detected  $F_I^\textrm{$\ast$} = 0.16\pm0.06~\mu$Jy,
i.e., 27$\pm$10\%  of the 
pulsar flux $F_I^\times$, which can be considered 
as an estimate for the contamination. 
This value is comparable with the 
3$\sigma$ uncertainty of
the flux measured in the pulsar's aperture
as well as with the difference 
between the $F_I^+$ and $F_I^\times$ fluxes 
(see Table~\ref{t:656_phot}). 
Notice, however, that some contribution to the flux at the
`{\large $\ast$}' position can also be provided by o3, so that
the real contamination from o2   
may be somewhat lower than estimated above. Moreover,
the assumption on symmetry of o2 
in the I band should be taken with caution
because, according to Fig.~\ref{f:656_bta},
the brightness distribution 
of o2 (and, perhaps, of o3)
in this filter appears to be different  
from those in the other bands.   

The exposure in the B band was too short to estimate the contamination.
Since $F_B^+ - F_B^\times \sim 0.1$~$\mu$Jy
is comparable to $F_I^+ - F_I^\times$, we can expect that
the pulsar's B flux can also be contaminated by o2.
Although this object is not seen in the 
adjacent F430W filter and in the other \hst\ UV bands,
its southern fragment is seen in the much 
wider F130LP filter whose band-pass strongly 
overlaps with that of the B filter.

Given the large uncertainty of the contamination,
it is very difficult to obtain accurate values
of the B, R and, particularly, I fluxes. 
Since the $F^+$ fluxes, measured by \cite{Kurt}, are obviously overestimated,
we can choose either $F^\times$ or $F^\times -F^\textrm{$\ast$}$ as
most probable values. Since $F_R^\textrm{$\ast$} \ll F_R^\times$, and
$F_B^\textrm{$\ast$}$ is difficult to measure, we choose
$F_R^\times$ and $F_B^\times$ 
as ``best-guess values''. As for the I flux, we present
both the $F_I^\times$ and $F_I^\times - F_I^\textrm{$\ast$}$ values
in Table~\ref{t:656_final}, where   
we collected all
optical data for the \psh.
\label{Ivalues}
Only observations 
with spatial resolution better than 1\arcsec\ can improve 
the accuracy of the flux measurements in the BRI bands.


\section{Discussion}

\label{s:discussion}

\subsection{Spectrum of \psh}

\begin{figure*}
\includegraphics[width=180mm,clip]{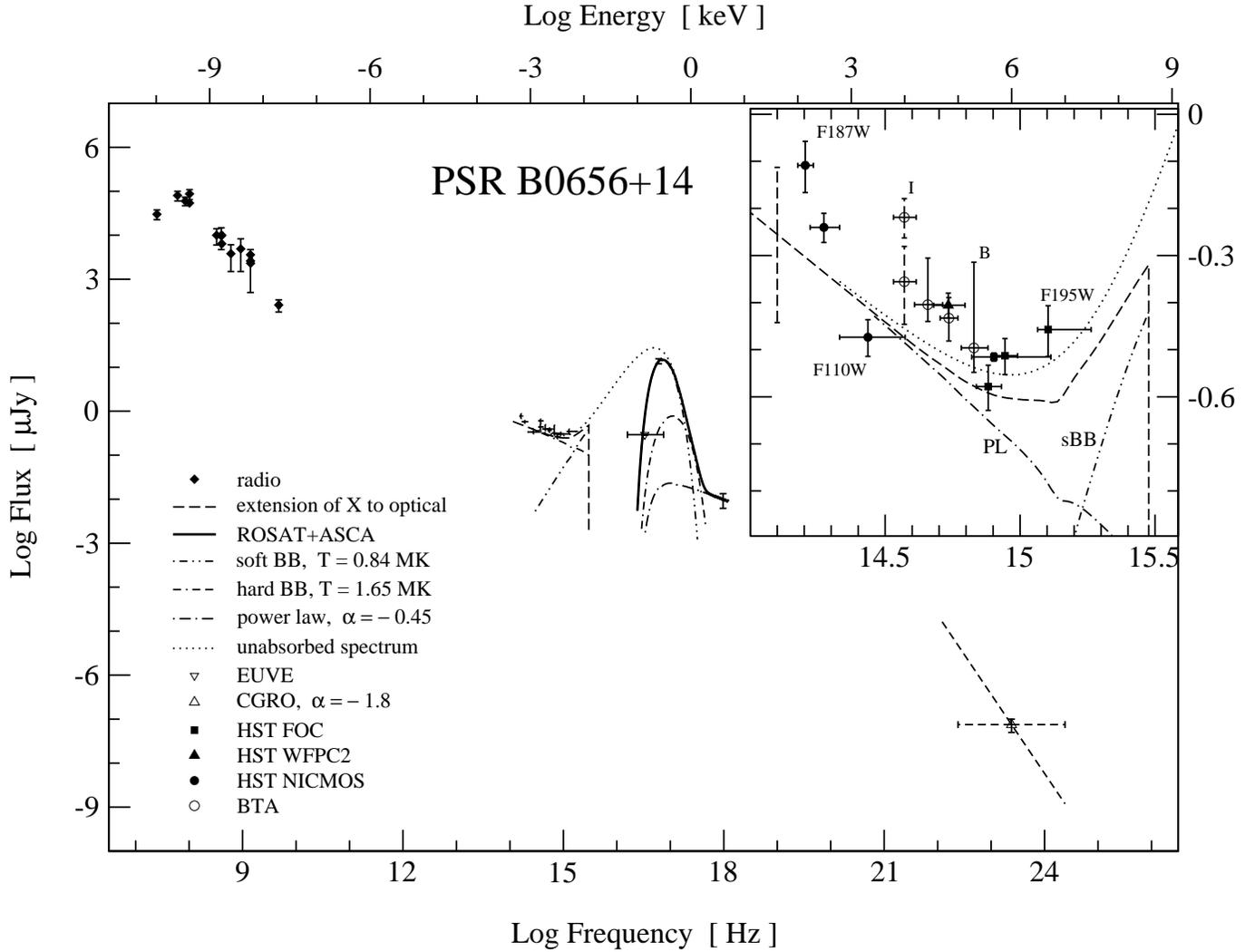}
\caption{
Broadband spectrum of \psh. The inset blows up
the optical range of the spectrum.
}
\label{f:sp_broad}
\end{figure*}

In Fig.~\ref{f:sp_broad} we 
combined our optical data with the available 
multiwavelength data on \psh\
(\cite{Mal}; \cite{PSC}; \cite{PWC}; \cite{EUVE}; \cite{r96}; \cite{zph00}).
The spectrum generally fades towards 
higher frequencies, showing a hump of thermal radiation
in soft X-rays.
  
In  X-rays, the {\sl EUVE}, {\sl ROSAT} and {\sl ASCA} 
data fit well with a three-component 
spectral model (\cite{zph00}). The model consists of a soft 
blackbody component describing thermal emission 
from the entire NS surface with a radius 
$R_s = 13.5\pm1.0$ km and temperature $T_s = (0.84\pm0.03)\times10^6$~K,  
a hard blackbody component representing thermal emission from 
pulsar hot polar caps with $T_h = (1.65\pm0.19)\times10^6$ K and 
$R_h = 0.9\pm0.5$ km, and a power-law non-thermal component, 
presumably formed in the NS magnetosphere, 
$f_E = C E^\alpha$ phot cm$^{-2}$ s$^{-1}$ keV$^{-1}$, 
where $E$ is photon energy in keV, $\alpha = -1.45\pm0.26$ 
and $C = (2.8\pm0.1)\times10^{-5}$. 
The radii $R_s$ and $R_h$ are related to the pulsar distance 
$d = 500$ pc; column density of the interstellar matter 
towards the pulsar is $n_H = (1.42\pm0.06)\times10^{20}$ cm$^{-2}$. 
The sum of the three components is displayed with the bold line  
in Fig.~\ref{f:sp_broad}; the unabsorbed 
spectrum is shown with the dotted line;
the dashed line represents an absorbed
extension of this three-component
model to the optical domain;
the dot-dashed lines represent contributions of 
different components of the fit. The soft thermal component 
from the entire surface of the cooling NS dominates 
in soft X-rays and in the {\sl EUVE} band.
The non-thermal component is responsible for the high energy 
tail in X-rays. It is also apparently detected  
in $\gamma$-rays, but its slope is highly uncertain  
and likely different from that in X-rays.

The IR-optical-UV range is blown up 
in the inset of Fig.~\ref{f:sp_broad}. 
It is seen that the observed broad-band spectrum 
is consistent, within the uncertainties, 
with the continuation of the X-ray fit to the optical range.
A typical uncertainty of the extrapolated flux is indicated 
by the dashed error-bar on the left side of the inset. 
The power-law component becomes dominating in the optical-IR range.
This suggests that a single mechanism of non-thermal radiation 
may operate in a broad frequency range from IR to hard X-rays. 
The slope of the non-thermal component
is less steep than it was deduced  by \cite{Kurt} 
and \cite{PWC} based on the previous optical-UV data.
The overall optical spectrum has a shallow minimum at 
$\nu \approx 7.4 \times 10^{14}$ Hz ($\lambda \approx 4000$ \AA). 
Above this crossover frequency, the Rayleigh-Jeans part, 
$F_\nu \propto \nu^2$, of the soft thermal component 
apparently dominates in the pulsar spectrum (\cite{PWC}).

From the new BTA and NICMOS data we cannot 
exclude nonmonotonous behavior of the broad-band IR-optical spectrum, 
which might indicate the presence of unresolved 
spectral  features. There are an apparent excess in B
(\cite{Kurt}; \cite{PWC}), an increase towards RI, and a dip in F110W, 
accompanied by  a steep increase towards the IR frequencies. 
The BRI excesses could be partially due to contamination of the pulsar flux 
by the nearby extended object o2.  
In the R and B filters they could be caused 
by H$_\alpha$ and H$_\beta$ emission from an unresolved pulsar-wind 
nebula (PWN) around the pulsar (e.g., \cite{c96}). 
For typical ISM temperatures and densities, 
the pulsar's velocity, $v_{\rm p} \ga 100 d_{500}$ km~s$^{-1}$ 
is supersonic, so that the PWN would be a bow shock 
at a stand-off distance 
$R_w \approx (\dot{E}/4\pi c\rho v_{\rm p}^2)^{1/2}=0.0013\dot{E}_{33}^{1/2}
n^{-1/2} v_{{\rm p},7}^{-1}$ pc, where 
$n=\rho/m_H$ is the baryon number density
of the ambient medium, $v_{{\rm p},7}=v_{\rm p}/(100~{\rm km}~{\rm s}^{-1})$.
For instance, such a shock,
with an H$_\alpha$ flux $f_{H_\alpha}^{(0437)}
 = 2.5\times 10^{-3}$ photons cm$^{-2}$ s$^{-1}$,
was observed at $7''$ ($R_w=0.006\, d_{180}$ pc 
for $\dot E^{(0437)} = 1.2\times10^{34}$ erg s$^{-1}$)
from the millisecond pulsar J0437--4715 (\cite{bell95}).
Since $f_{H_\alpha}\propto\dot{E} v_{\rm p} X d^{-2}$,
where $X$ is the fraction of neutral hydrogen atoms,  
the expected H$_\alpha$ flux from a bow shock at $\sim 2''$--$3''$
from \psh\ can be estimated by simple scaling:
$f_{H_\alpha}^{(0656)} = 1.0\times 10^{-3} X^{(0656)}/X^{(0437)}$
photons cm$^{-2}$ s$^{-1}$, for $v_{\rm p}^{(0437)}=v_{\rm p}^{(0656)}=
100$ km~s$^{-1}$, $d^{(0656)}=500$ pc, $d^{(0437)}=180$ pc.
This corresponds to an R-band energy flux 
$F_R^{(0656)} \approx 2\, X^{(0656)}/X^{(0437)}$
$\mu$Jy from the whole H$_\alpha$ emitting region. 
The contribution of this flux in the observed R flux would be very substantial
at $X^{(0656)}=X^{(0437)}$. However, 
\psh\ is much younger than J0437--4715,
and its hot NS creates an ionization zone of a radius $R_s\sim 0.3$--0.5 pc
($\sim 2'$--$3'$) around the pulsar (e.g., \cite{Schwarts}).
Since $X^{(0656)}$ at $r=R_w$ ($\ll R_s$) is expected to be orders of
magnitude smaller than $X^{(0437)}$,
the contribution of the bow-shock H$_\alpha$ emission in $F_R$ should be
negligible. 
The estimated angular size of the ionization zone is much larger than our apertures,
its emission is hidden in background and cannot contribute to the measured
point source flux.

\begin{figure}[t]
\includegraphics[width=85mm,clip]{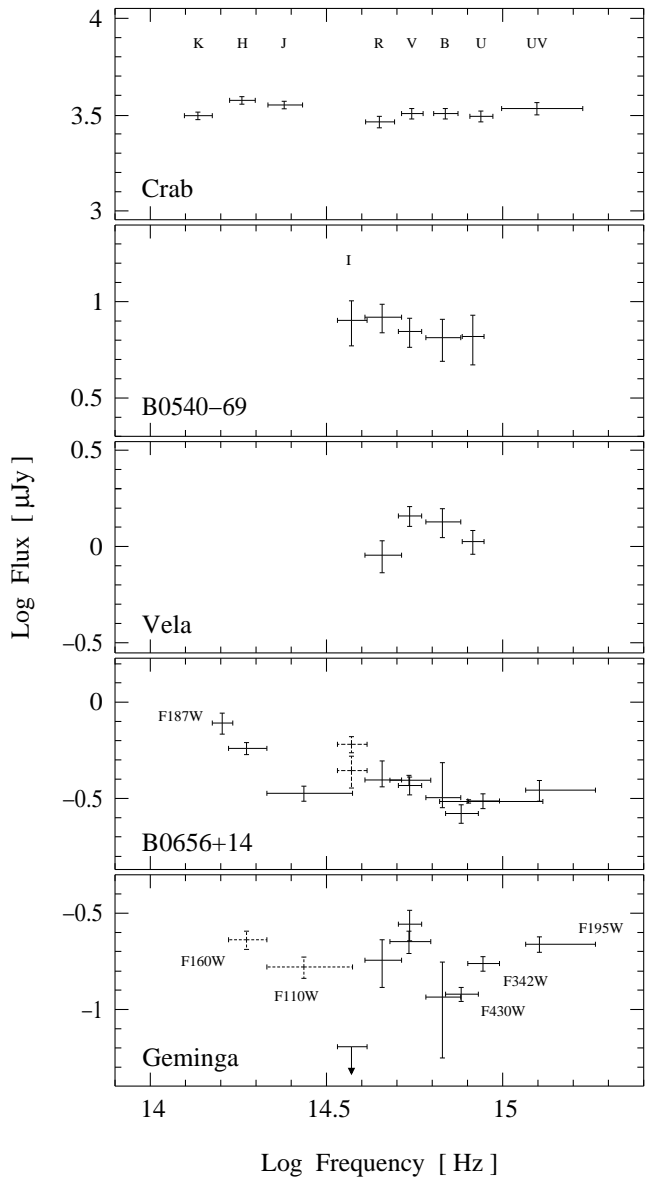}
\vspace{3mm}
\caption{
Illustration of spectral evolution of pulsars with age,
from the younger Crab pulsar in the {\it top\/} panel
to the older Geminga pulsar in the {\it bottom\/}.
}
\label{f:sp_evol}
\end{figure}

A similar excess, albeit in the V band,
is also present in the broad-band spectrum of the other
well-studied middle-aged pulsar \object{Geminga}.  
Its  possible explanations have been suggested 
by \cite{bcmeb96} and \cite{mhs98} based on various 
combinations of the synchrotron or ion cyclotron
emission/absorption in an upper atmosphere 
or low magnetosphere of the NS.  
In Fig.~\ref{f:sp_evol} we plot the optical-UV fluxes 
(from \cite{mcb98}) of Geminga together with our estimates 
of the NICMOS F110W and F160W fluxes detected by \cite{Harlow}. 
We also show the available dereddened color 
spectra of  the Crab pulsar (\cite{Perc}; \cite{efrmkp97}), 
PSR B0540$-$69
(\cite{mpb87}), the Vela pulsar (\cite{nmcb97}), and
\psh\ (see Table~\ref{t:656_final}).
We neglected the extinction for \psh\ and Geminga,  
which gives rather small corrections for these nearby pulsars.
It is seen that the broad-band spectra 
of \psh\ and Geminga are 
qualitatively similar to each other. 
This suggests that the unresolved  spectral features 
associated with the apparent optical excesses in their spectra 
could be of the same nature.
On the other hand, the spectra of the middle-aged pulsars
appear to be different from the flat and featureless spectra of
the young Crab and B0540$-$69 pulsars, which is evidence of
evolution of optical emission with pulsar age.
To confirm the difference
and reveal its nature, spectroscopic observations of the
middle-aged pulsars are needed.

\begin{table}[t]
\caption{Pulsar fluxes in the optical-UV bands. 
}
\label{t:656_final}
\begin{tabular}{c|ccc}
\hline\hline
Band   & $M^a$                & Flux                  & References$^b$ \widerul \\
       &                      & $\mu$Jy                & data / measured by \\
\hline
B      & 25.25$^{+29}_{-15}$  & 0.32$^{+10}_{-04}$     & tw\widerul \\
V      & 24.97(12)            & 0.369(39)              & tw   \\
R      & 24.72$^{+25}_{-09}$  & 0.39$^{+10}_{-03}$     & tw    \\
I$^c$  & \twolines{23.99(11)}{24.33(23)} 
       & \twolines{0.604(58)}{0.441(83)}               & tw   \\
\hline
F555W  & 24.90(07)            & 0.393(24)              & M97a    / tw \\
F110W  & 24.38(10)            & 0.336(30)              & H98    / tw \\
F160W  & 23.22(08)            & 0.575(41)              & H98    / tw \\
F187W  & 22.63(13)            & 0.779(97)              & H98    / tw \\
F430W  & -                    & 0.264(29)              & \cite{PWC}   \\
F342W  & -                    & 0.307(27)              & \cite{PWC}   \\
F195W  & -                    & 0.349(43)              & \cite{PWC}   \\
F130LP & -                    & 0.305(06)              & \cite{PSC}   \\
\hline
\end{tabular}
\renewcommand{\arraystretch}{0.8}
\begin{tabular}{cl}
& \\
$^a$ & Johnson-Cousins (for the BTA images) or Vega system \\
     & (for the \hst\ images) magnitude \\
$^b$ & M97a -- \cite{Mign}, H98 -- \cite{Harlow}, \\
     & tw -- this work \\
$^c$ & See explanations in Sect.~\ref{Ivalues} \\
\end{tabular}
\end{table}

\subsection{Nearby objects}  

{\it Object o5.}
Faintness of the stellar object o5 suggests that it might be a main sequence star  
located near the edge of our galaxy.   
Taking into account the total interstellar galactic extinction towards 
the pulsar field ($l=201\fdg1$, $b=8\fdg26$;  $A_B=0.420$, $A_V=0.323$, 
$A_R=0.260$, $A_I=0.189$, $A_J=0.088$, $A_H=0.056$, 
$A_K=0.036$ -- see \cite{Schl98}), we obtain  the dereddened
color indices of o5:  $M_{F555W}^\mathrm{Vega}-R=0.85(9)$
and $R-I=0.62(12)$. This would be compatible 
with a K6V main sequence star (\cite{Bes90}). 
However, typical absolute magnitude of the K6V stars 
are $M_V$ = 8\fm2--8\fm5. This yields 
a distance range of 20--23 kpc, 
incompatible with the Galaxy size.
Therefore, it is more likely that o5 is a white dwarf  
with $T\sim 4000$ K and $M_V\sim16^m$,
at a distance of  $\sim 600$ pc. However, in the absence 
of spectral information one cannot rule out also 
that this is a quasar.

{\it Extended objects.}
There were unsuccessful attempts to search for a PWN around
\psh\ on scales $\ga 10$\arcsec\
in radio (\cite{fs97}) and X-rays (\cite{Beck_X}).
However, an expected PWN size, 
$\sim [\dot{E}/(4\pi cp_0)]^{1/2}$,
where $p_0$ is the pressure of the ambient medium, can be smaller
than $10''$, and, in principle, the object o2 might be a PWN candidate.
The morphology of o2 looks like that of a faint 
spiral galaxy seen edge-on.
However, its size, $\sim$~3\arcsec~$\times$~0\farcs7,
is larger than the mean size,
$\sim$~1\arcsec~$\times$~0\farcs5,
of galaxies of close magnitudes,
$m^{AB}_{F160W} \simeq 22$, in the South
Hubble Deep Field NICMOS catalog\footnote{See
nicmos\_v1generic.cat, presented at \newline
\href{http://www.stsci.edu/ftp/science/hdfsouth/hdfs.html}
{http://www.stsci.edu/ftp/science/hdfsouth/hdfs.html}.}
taken with the exposure 4992~s close to
the pulsar field exposure 5098~s in the F160W band.
The broad-band spectrum of o2
is close to a power law with a photon index of about $-3.4$.
If it were a nebula powered by relativistic particles ejected
by the pulsar, 
its observed flux could be provided by synchrotron radiation
from a region with approximately the same parameters
as those estimated by Hester et al.~(1995)
 for the optical structures (wisps, knots)
in the synchrotron nebula around the much more energetic Crab pulsar
(e.g., from a volume $V \sim 2 \times 10^{48}$~ cm$^3$
with a magnetic field $B\sim 3 \times 10^{-3}$~G
and a steep power-law distribution 
of relativistic electrons with a slope $p=5.8$,
number density $n_e\sim 3 \times 10^{-5}$~cm$^{-3}$,
and minimum Lorentz factor $\gamma_{\rm min}\sim 10^4$).
However, the morphology of o2 looks very unnatural for a PWN,
particularly with account of the pulsar's proper motion (see Fig.~2),
the slope of the power-law spectrum is too steep compared with
other known synchrotron sources,
and \psh\ can hardly supply relativistic particles at a rate needed
to support the amount required.
Thus, we conclude that o2 is most likely a faint background galaxy,
as are the other extended objects o3 and o4.

Deep subarcsecond optical, radio and X-ray observations of the \psh\  
field are needed to search for possible unresolved  structures associated  
with activity of \psh. In this respect, it would be useful to 
confirm and investigate a knot-like structure seen
at a $3\sigma$ level above the background at $\sim$0\farcs04
from the pulsar position
in the FOC/F342W image\footnote{We are grateful 
to R.~Mignani who checked this 
independently.}. 


\begin{acknowledgements} 

Partial support for this work was provided by grant 1.2.6.4
of Program "Astronomia", by INTAS (grant 96--0542),
RFBR (grant 99-02-18099), and by NASA through grant
number GO-07836-01-96A from the Space Telescope Science Institute,
which is operated by the Association of Universities for Research
in Astronomy, Incorporated, under NASA contract NAS5-26555.
We are grateful to Roberto Mignani for the \hst/WFPC2 data,
to Jason Harlow, Richard Hook, Andrew Fruchter,
Brian Monroe (NICMOS Helpdesk), and Lindsey Davis (IRAF Helpdesk) 
for the assistance, to Victorija Komarova for careful reading.

\end{acknowledgements}

\end{document}